%% file: main.tex
\documentclass{article}
\usepackage[utf8]{inputenc}
\usepackage{fixltx2e,fix-cm}
\usepackage{amssymb}
\usepackage{amsmath}
\usepackage{graphicx}
\usepackage{makeidx}
\usepackage{multicol}
\usepackage[title]{appendix}
\usepackage{macros/packages}
\usepackage{macros/statistics}
\usepackage{macros/editing}
\usepackage{macros/formatting}
\graphicspath{ {./images/} }

\newcommand{\EE}[2][]{\mathbb{E}_{#1}\left[#2\right]}
\newcommand{\dataset}{\mathcal{O}}
\newcommand{\Rlogo}{\protect\includegraphics[height=1.8ex,keepaspectratio]{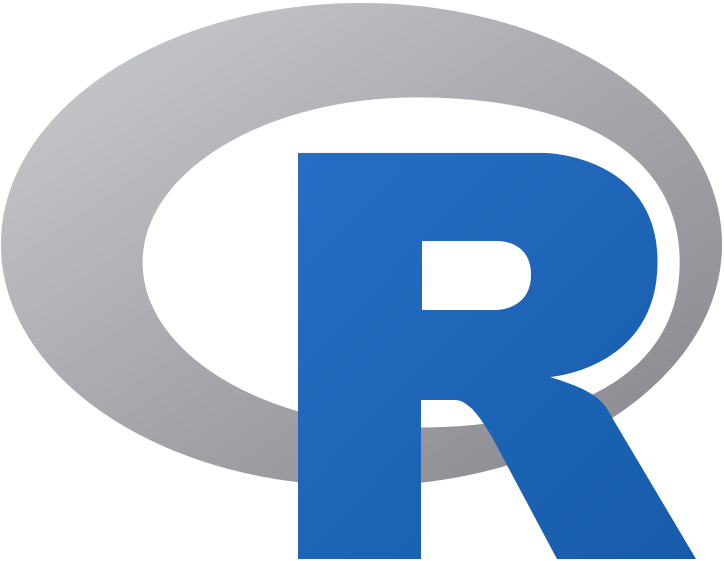}}

\usepackage{graphics}
\usepackage{nicematrix}
\definecolor{bg}{rgb}{0.95, 0.95, 0.96}
\definecolor{ponyolightyellow}{HTML}{ECD89D}
\definecolor{ponyomediumyellow}{HTML}{D8AF39}
\definecolor{ponyolightpink}{HTML}{F4E3D3}
\definecolor{ponyomediumpink}{HTML}{E8C4A2}
\definecolor{ponyolightblue}{HTML}{94C5CC}
\definecolor{ponyomediumblue}{HTML}{278B9A}

\usepackage{enotez}
\let\footnote=\endnote
\DeclareInstance{enotez-list}{custom}{paragraph}{
heading = \section{#1}\label{sec:notes}
}

\title{Treatment Heterogeneity for Survival Outcomes}
\author[1]{Yizhe Xu\thanks{corresponding to yizhex@stanford.edu}}
\author[2]{Nikolaos Ignatiadis}
\author[3]{Erik Sverdrup}
\author[1]{Scott Fleming}
\author[3]{Stefan Wager}
\author[1]{Nigam H. Shah}
\affil[1]{Stanford Center for Biomedical Informatics Research, Stanford University, Stanford, CA}
\affil[2]{Department of Statistics, Stanford University}
\affil[3]{Graduate School of Business, Stanford University}

\begin{document}

\maketitle

\begin{abstract}
    Estimation of conditional average treatment effects (CATEs) plays an essential role in modern medicine by informing treatment decision-making at a patient level. Several metalearners have been proposed recently to estimate CATEs in an effective and flexible way by re-purposing predictive machine learning models for causal estimation.  In this chapter, we summarize the literature on metalearners and provide concrete guidance for their application for treatment heterogeneity estimation from randomized controlled trials' data with survival outcomes. The guidance we provide is supported by a comprehensive simulation study in which we vary the complexity of the underlying baseline risk and CATE functions, the magnitude of the heterogeneity in the treatment effect, the censoring mechanism, and the balance in treatment assignment. To demonstrate the applicability of our findings, we reanalyze the data from the Systolic Blood Pressure Intervention Trial (SPRINT) and the Action to Control Cardiovascular Risk in Diabetes (ACCORD) study. While recent literature reports the existence of heterogeneous effects of intensive blood pressure treatment with multiple treatment effect modifiers, our results suggest that many of these modifiers may be spurious discoveries. This chapter is accompanied by \mintinline{R}{survlearners}, an \Rlogo~package that provides well-documented implementations of the CATE estimation strategies described in this work, to allow easy use of our recommendations as well as reproduction of our numerical study.
\end{abstract}

\section{Introduction}

Healthcare decisions are commonly informed by a combination of \emph{average treatment effects} (ATEs) from Randomized Controlled Trials (RCTs), which do not account for fine-grained patient heterogeneity and \emph{risk stratification} that identify those at the highest need for an intervention. For example, physicians assess patients' ten-year Atherosclerotic cardiovascular disease (ASCVD) risk based on baseline covariates (such as age, blood pressure, and cholesterol levels) using the Pooled Cohort Equations (PCEs)~\citep{goff2014} and initiate statin treatment based on clinical practice guidelines~\citep{arnett2019}. Prioritizing treatment to patients who are at higher risk is a sensible starting point; however, it may be suboptimal when baseline risk is not an appropriate surrogate for treatment effects. For example, a young patient with only one risk factor of elevated cholesterol level probably has a lower ASCVD risk than an older subject who has a normal cholesterol level but multiple other risk factors, such as high blood pressure and smoking, but the younger patient may benefit more from statin which reduces cholesterol~\citep{soran2015cholesterol, thanassoulis2017benefit}. Therefore, a key question in delivering personalized care remains: \emph{How} should we treat each \emph{individual} patient?

Consequently, a key challenge in enabling precision medicine is to go beyond estimation of ATEs and risk stratification, and account for the varied patient response to treatment depending on factors such as patient characteristics, baseline risk, and sensitiveness to treatment~\citep{Kravitz2004}.  Such heterogeneity in treatment effects (HTE) may be summarized by  conditional average treatment effects (CATEs) as a function of subject-level baseline covariates. CATE estimation however is a difficult statistical task. Estimating CATEs essentially requires estimating interactions of treatment and baseline covariates, and such interaction effects are often small compared to main effects that drive baseline risk. Conventional one-variable-at-a-time and subgroup analyses may produce false-positive results due to multiple testing and false-negative results due to insufficient power, i.e., small sample size in subgroups \citep{Wang2007}. 

As we discuss more in the related work section below, a promising approach towards CATE estimation that addresses some of the above challenges is the development of estimation strategies that use flexible machine learning methods. A particularly convenient class of such methods are called metalearners~\citep{Kunzel2019,Nie2020}: The premise is that one can decompose the CATE estimation task into well-understood machine learning tasks (as would be conducted, e.g., for risk stratification). Statistical and domain expertise in risk modeling can thus be repurposed for a causal task, namely estimation of heterogeneous treatment effects.

In this chapter, we build on the metalearner literature and provide concrete guidance for their usage in estimating treatment effect heterogeneity from randomized controlled trial data with right-censored survival outcomes. Our contributions are as follows: 1) We provide an accessible summary of the mathematical underpinning of five popular metalearners (S-, T-, X-, M-, and R-learners) when combined with two popular machine learning strategies (Lasso and generalized random forests). While the described metalearners have been developed for uncensored continuous or binary data, we explain how they may be adapted to the survival setting through inverse probability of censoring weighting; 2) We provide \Rlogo~code in a package called \mintinline{R}{survlearners}~\citep{survlearners} that demonstrates exactly how these methods may be implemented in practice and describe how machine learning models (e.g., risk models) are leveraged by each metalearner; 3) We conduct a comprehensive simulation study of the above CATE estimation strategies using several data generation processes (DGPs) in which we systematically vary the complexity of the baseline risk function, the complexity of the CATE function, the magnitude of the heterogeneity in treatment effects, the censoring mechanism, and the imbalance of treated and control units; 4) Based on the results of the simulation study, we summarize considerations that matter in choosing and applying metalearners and machine learning models for HTE estimation; 5) We apply our findings as a case study of HTE estimation on the Systolic Blood Pressure Intervention Trial (SPRINT) \citep{Wright2015} and the Action to Control Cardiovascular Risk in Diabetes (ACCORD) trial \citep{Cushman2010}.  

\subsection{Related work}

Prior works have introduced metalearners for HTE estimation~\citep{Athey2016, Kunzel2019,Nie2020}, and provided comprehensive tutorials and simulation benchmarks that explicate their usage~\citep{caron2022}. Further works have conducted comprehensive  simulation studies that compare both metalearners, as well as other machine learning approaches to HTE estimation~\citep{Powers2017, Wendling2018, Feller2019, saito2019doubly,  curth2021nonparametric, curth2021really,jacob2021cate, knaus2021machine, ling2021heterogeneous, Naghi2021Finite, zhang2021unified, okasa2022meta}. More broadly, there have been substantial advances in nonparametric methodology for estimating HTEs~\citep{Athey2019, hahn2020bayesian, Hill2011, Foster2020, wager2018estimation, Yang2021} and in theoretical understanding thereof, e.g., with respect to minimax rates of estimation~\citep{Gao2020,kennedy2020optimal, kennedy2022minimax}. However, these studies are all conducted under a non-survival setting, while researchers in healthcare are often interested in right-censored, time-to-event outcomes. Our chapter fills this gap and provides a tutorial and benchmark of metalearners adapted to the survival analysis setting.

Several researchers \citet{jaroszewicz2014uplift, zhang2017mining, henderson2020individualized, tabib2020non, zhu2020targeted,  cui2021estimating, chapfuwa2021enabling, curth2021survite, hu2021epidemiology} have developed machine learning based methods for estimating HTEs with data that have time-to-event outcomes (often called \textit{survival data}). \citet{hu2021estimating} conduct a comprehensive simulation study of the T-learner (one of the metalearners described in this work) combined with different predictive models for survival data.~\citet{duan2019clinical} reanalyze the SPRINT trial and in doing so extend the X-learner (a metalearner proposed by~\citet{Kunzel2019}) to the survival setting through inverse probability of censoring weights. However, we are not aware of a previous comprehensive tutorial and benchmark that demonstrates how to use a range of popular metalearners with survival data. Furthermore, while much of the related literature focuses on the more general setting with observational data under unconfoundedness, we focus exclusively on RCTs. In doing so, we hope to clarify the essential difficulties of HTE estimation~\citep{hoogland2021tutorial} in the absence of confounding beyond censoring bias. We hope that our work will contribute to the principled estimation of HTEs from RCTs in medicine.

\subsection{The PATH Statement}
\label{subsec:path}

The Predictive Approaches to Treatment effect Heterogeneity (PATH) Statement provides guidance for estimating and reporting treatment effect heterogeneity in clinical RCTs and was developed by a panel of multidisciplinary experts~\citep{kent2020predictive}. The PATH statement identifies two main approaches to predictive HTE estimation: 
%\begin{enumerate}[leftmargin=*, label={}]

\emph{``PATH risk modeling:''} A predictive risk model is identified (or developed) and HTEs are estimated as a function of predicted risk in the RCT data.

\emph{``PATH effect modeling:''} A predictive model is developed for the outcome of interest with predictors that include the risk predictors, the treatment assignment, as well as interaction terms.

Our work is complementary to the PATH statement and provides further methodological guidance for HTE estimation in RCTs with survival outcomes. Concretely, the ``PATH effect modeling'' approach coincides with the ``S-learner'', a metalearner proposed in the machine learning literature and described in Section~\ref{subsec:slearner}. Our numerical results and literature review confirm the caveats of effect modeling described in the PATH Statement. Other metalearners may be preferable in settings where those caveats apply. The PATH guidelines for ``PATH risk modeling'' approach are important for all the metalearners considered in this work. These guidelines advocate for the use of a parsimonious set of predictors for HTEs. A risk score (developed previously, or blinded to treatment assignment) is an important predictor for HTEs that is justified both mathematically and  by clinical experience. In this work, by using modern machine learning and regularization techniques, we allow for the possibility that $X_i$ may be high-dimensional. We do not provide guidance on how to choose predictors $X_i$ ($X_i$ may include e.g., risk scores previously developed, but also additional  predictors determined based on domain expertise) but describe methods for efficiently using any such HTE predictors chosen by the analyst. 

\subsection{Outline}

The outline of this chapter is as follows. We define the CATE estimation problem in Section \ref{sec:setup}. In Section~\ref{sec:metalearners} we provide a brief tutorial on the use of metalearners and machine learning for estimating treatment heterogeneity in randomized controlled trials with right-censored, time-to-event data. We describe our simulation study in Section~\ref{sec:sim} and summarize main takeaways in Section~\ref{sec:sim_results}. We then present a case study on SPRINT and ACCORD in Section \ref{sec:sprint}. Finally, we conclude with a discussion and future extensions in Section~\ref{sec:dis}.

\section{Problem Setup, Notation, and Assumptions}\label{sec:setup}

We discuss the problem of CATE estimation under the potential outcome framework in causal inference \citep{Neyman1923, Rubin1974}. Consider an RCT that provides $N$ independent and identically distributed subject-level data, indexed by $i \in \mathcal{I} = \{1, 2, \dotsc, N\}$. Patient baseline covariates are denoted by $X_i = (X_{i1},\dotsc,X_{id}) \in \mathbb{R}^d$ and the treatment variable is $W_i \in \{0,1\}$. Let $T_i(w)$ be the potential survival time if subject $i$ had been assigned the treatment $w$ and $C_i$ be the censoring time. Then, $T_i = W_i T_i(1)+ (1-W_i)T_i(0)$, and the observed follow-up time is $U_i=\mathrm{min}(T_i, C_i)$ with the corresponding non-censoring indicator $\Delta_i = \ind{T_i \leq C_i}$. Our observed dataset is $\dataset = \{(X_i, W_i, U_i, \Delta_i): i \in \mathcal{I}\}$. We also reserve the notation $\mathcal{I}_0 = \{ i \in \mathcal{I}: W_i=0\}$ for the index set of control units and $\dataset_0 = \{(X_i, 0, U_i, \Delta_i): i \in \mathcal{I}_0\}$ for the control dataset. We define $\mathcal{I}_1$ and $\dataset_1$ analogously for the treated units. Occasionally we will explain concepts using \Rlogo~pseudocode in which case we will refer to the observed data as \mintinline{R}{X} ($N \times d$ matrix) and \mintinline{R}{W}, \mintinline{R}{U}, and \mintinline{R}{D} (length $N$ vectors with $i$-th entry equal to $W_i$, $U_i$, resp. $\Delta_i$). 

The Conditional Average Treatment Effect (CATE) for the probability of survival beyond a pre-specified time $t_0$ is:
\begin{equation}
\label{eq:CATE}
\tau(x) = \EE{Y_i(1)-Y_i(0) \mid X_i=x} ,
\end{equation}
where $Y_i(w) = \ind{T_i(w) > t_0}$ is the indicator of survival beyond time $t_0$. We also write $Y_i = Y_i(W_i)$.

To ensure the CATE is identifiable, we make the following assumptions \citep{ROSENBAUM1983}; some of which already appeared implicitly in the notations above:
\begin{itemize}[leftmargin=*, label={}]
\item \emph{Consistency}: The observed survival time in real data is the same as the potential outcome under the actual treatment assignment, i.e., $T_i=T_i(W_i)$.
\item \emph{Randomized controlled trial (RCT)}: The treatment assignment is randomized such that $W_i$ is independent of $(X_i, T_i(1), T_i(0))$, i.e., $W_i \independent (X_i, T_i(1), T_i(0))$ and $\mathbb P[W_i=1] = e$ with known $0<e<1$.
\item \emph{Noninformative censoring}: Censoring is independent of survival time conditional on treatment assignment and covariates, i.e., $C_i \independent T_i\,\mid\,X_i, W_i$.
\item \emph{Positivity}: There exists subjects who are at risk beyond the time horizon $t_0$, i.e., $\mathbb{P}[C_i > t_0\,\mid\,X_i, W_i] \geq \eta_{C}$ for some $\eta_{C}>0$. 
\end{itemize} 

\begin{remark}[Observational studies]
\label{remark:observational}
As mentioned above, throughout this manuscript we focus our attention on randomized controlled trials so as to provide a comprehensive discussion of issues involved in the estimation of HTEs in the absence of confounding beyond censoring bias. Nevertheless, conceptually, the metalearners we discuss are also applicable in the setting of observational studies under unconfoundedness. Concretely, we may replace the \emph{RCT} assumption by the following two assumptions
\begin{itemize}[leftmargin=*, label={}]
\item \emph{Unconfoundedness:} The potential survival times are independent of the treatment assignment $W_i$ conditionally on baseline covariates, that is, $W_i \independent (T_i(1), T_i(0)) | X_i$.
\item \emph{Overlap:}  There exists $\eta \in (0,1)$ such that the propensity score $e(x) = \mathbb P[W_i=1 |X_i=x]$ satisfies $\eta \leq e(x) \leq 1-\eta$ for all $x$ in the support of $X_i$.
\end{itemize}
We refer the interested readers to the original manuscripts introducing the different metalearners for an explanation of their application to observational studies. In short, for the methods we describe, it suffices to replace the treatment probability $e$ (whenever it is used by a method) by $\widehat{e}(X_i)$ with $\widehat{e}(\cdot)$ is an estimate of the propensity score $e(\cdot) = \mathbb P[W_i=1 |X_i=\cdot]$. The statistical consequences of estimation error in $\widehat{e}(\cdot)$, as well as ways to make estimators robust to this error, are discussed further in \citet{Nie2020} and \citet{Foster2020}.
\end{remark}

\section{Metalearners}
\label{sec:metalearners}
Metalearners are specific meta-algorithms that leverage predictive models to solve the causal task of estimating treatment heterogeneity. Metalearners are motivated by the observation that predictive models are applied ubiquitously, that we have a good understanding about fitting models with strong out-of-bag predictive performance, and we know how to evaluate predictive models~\citep{hastie2009elements, van2011dynamic,harrell2015regression}. Metalearners repurpose this expertise to power the effective estimation of HTEs (a task which is less well-understood). Proposed metalearners build upon different predictive tasks and also combine these predictive models in distinct ways to estimate HTEs. In this Section we seek to provide a short, but instructive, introduction to commonly used metalearners in the context of CATE estimation~\eqref{eq:CATE} with survival data.

To emphasize the flexibility of metalearners in leveraging predictive models, we abstract away the concrete choice of predictive model for each task by introducing the notation
\begin{equation}
\label{eq:risk_prediction}
\mathcal{M}\left(\widetilde{Y} \sim \widetilde{X};\; \widetilde{\dataset}, [\widetilde{K}]\right),
\end{equation}
to denote a generic prediction model that predicts $\tilde{Y}$ as a function of covariates $\widetilde{X}$ based on the dataset $\widetilde{\dataset}$ with (optional) sample weights $\widetilde{K}$. Note that by default we assume that $\widetilde{K}_i =1$ for all $i$, in which case we omit $\widetilde{K}$  from the notation and write $\mathcal{M}(\widetilde{Y} \sim \widetilde{X};\; \widetilde{\dataset})$. It will also be convenient to introduce notation for predictive models that give out-of-bag (oob)\footnote{We use the term out-of-bag loosely also using it to refer to out-of-sample or out-of-fold predictions.} predictions of $\widetilde{Y}$:
\begin{equation}
\label{eq:risk_prediction_oob}
\mathcal{M}^{\text{oob}}\left(\widetilde{Y} \sim \widetilde{X};\; \widetilde{\dataset}, [\widetilde{K}]\right).
\end{equation}
Below we will describe the high-level idea of each metalearner, followed by concrete examples for possible choices of $\mathcal{M}$. We first describe two metalearners---the S- and T-learners---for which we only need to predict the probability that $\{T_i > t_0\}$ as a function of certain covariates $\widetilde{X}_i$. We refer to this task as risk modeling, since
$$\mathbb P[T_i > t_0 \mid \widetilde{X}_i] \, =\, 1-\mathbb P[T_i \leq t_0 \mid \widetilde{X}_i]\, =\, 1- \text{Risk}(\widetilde{X}_i).$$
Standards for transparent reporting of predictive risk models~\citep{collins2015transparent} should be adhered to in the course of developing such models within the context of metalearners. 

We emphasize that risk modeling, as defined in our work, has subtle differences from the ``PATH risk modeling'' approach (Subsection~\ref{subsec:path} and \citet{kent2020predictive})---to avoid any confusion we quote the latter throughout. One key difference is that, for us, risk models are always assumed to be developed on the training dataset $\dataset$ and not in a previous cohort ($\widetilde{\dataset}$ and $\dataset$ are the general and case-specific notations of a data set, respectively. This distinction also applies to the $\widetilde{Y}$, $\widetilde{X}$, and $\widetilde{K}$ in (\ref{eq:risk_prediction}) throughout the chapter). Nevertheless, as we already alluded to in Subsection~\ref{subsec:path}, the two frameworks are compatible with each other. We will further demonstrate their connection in our case-study reanalyzing the SPRINT RCT in Section ~\ref{sec:sprint}: There we use the well-established Pooled Cohort Equations (PCE)~\citep{goff2014} that predict the 10-year risk of a major ASCVD event as a function of predictors such as total cholesterol, blood pressure, age, etc. This gives us a new feature---the PCE score---which can then be used as a predictor for risk modeling within the SPRINT~\citep{Wright2015} cohort.

\subsection{S-learner: Modeling Risk as a Function of Baseline Covariates, Treatment Assignments, and their Interactions}
\label{subsec:slearner}
The ``simplest'' metalearner used in practice, called the  \emph{S-learner}, fits a \emph{single} risk model using baseline covariates and the treatment variable. The S-learner builds upon the observation that under our assumptions, the CATE in~\eqref{eq:CATE} may be written as:
\begin{equation}
\label{eq:slearner_identification}
\tau(x) =  \mu([x,1]) - \mu([x,0]) \text{ with } \mu([x,w]) = \mathbb E[Y_i \mid X_i=x, W_i=w].
\end{equation}
In the notation of~\eqref{eq:risk_prediction}, the S-learner proceeds as follows:
\begin{equation}
\label{eq:slearner}
\what{\mu}(\cdot) = \mathcal{M}(Y \sim [X, W];\; \dataset),\;\;\; \what{\tau}(x) = \what{\mu}([x, 1])-\what{\mu}([x, 0]).
\end{equation}
In words, the S-learner first learns a risk model $\what{\mu}(\cdot)$ as a function of $X_i$ and $W_i$, i.e., the treatment assignment $W_i$ is merely treated as ``just another covariate'' ~\citep{Hill2011, hahn2020bayesian}. Then, given baseline covariates $x$, the S-learner applies the fitted model to $[x,1]$ (that is, to the feature vector that appends $W_i=1$ to $x$) to impute the response of the treated outcomes. The fitted model is then applied to $[x, 0]$ to impute the response of the control outcomes, and finally the CATE is estimated as the difference thereof. 

In~\eqref{eq:slearner_identification}, any predictive model for $\mathbb P[T_i > t_0 \mid X_i, W_i]$ may be used, for example a random survival forest.

\begin{example}[S-learner with Random Survival Forest]
\label{example:grfS}
A concrete example of a fully nonparametric model \smash{$\mathcal{M}(Y \sim [X, W];\; \dataset)$}  is the random survival forest of \citet{ishwaran2008random}. The basic idea of the random survival forest, and more generally, of generalized random forests (GRF)~\citep{Athey2019} is the following: As in the traditional random forest of \citet{Breiman2001}, a collection of trees is grown. Each tree is grown based on a randomly drawn subsample of the training data and by recursively partitioning the covariate space. These trees are then used to adaptively weight~\citep{lin2006random} new test points. To be precise, let $\tilde{x}$ be a test covariate (e.g., $\tilde{x} = [x, w]$ for the S-learner), then the $i$-th data point in the training dataset is assigned weight \smash{$\alpha_i(\tilde{x}) =  \sum_{b=1}^{B}\mathbf{1}(\{\widetilde{X}_i \in L_b(\tilde{x})\})/(B|L_b(\tilde{x})|),$}
where $B$ is the total number of trees, $L_b(\tilde{x})$ is the set of all training examples $i$ such that $\widetilde{X}_i$ falls in the same leaf of the $b$-th tree as $\tilde{x}$, and $|\cdot|$ symbolizes the number of instances in a set. Then, given these weights $\alpha_i(\tilde{x})$, the Nelson-Aalen estimator with these weights is used to predict \smash{$\mathbb P[T_i > t_0 \mid \tilde{X}_i]$}. Using the random survival forest in the \Rlogo~package \mintinline{R}{grf}~\citep{grf}, the S-learner may be implemented as follows:
\begin{minted}[fontsize = \footnotesize,
               bgcolor = bg]{R}
library(grf)
m <- survival_forest(data.frame(X=X, W=W), U, D, prediction.type="Nelson-Aalen")
m1_x <- predict(m, data.frame(X=x, W=1), t0)$predictions
m0_x <- predict(m, data.frame(X=x, W=0), t0)$predictions
tau_x <- m1_x - m0_x
\end{minted}
In the first line, we load the \mintinline{R}{grf} package. In the second line, we fit the random survival forest based on the full dataset with augmented covariates for which we concatenate the baseline covariates \mintinline{R}{X} and treatment assignment vector \mintinline{R}{W} and with follow-up times \mintinline{R}{U} and event indicators \mintinline{R}{D}. Then, given test baseline covariates $\mintinline{R}{x}$, we impute (predict) the survival probability at $t_0$ (\mintinline{R}{t0} is a scalar that equals to $t_0$) using the fitted survival forest for the treated outcome (third line) and control outcome (fourth line), and finally we take their difference (fifth line).
\end{example}

The random survival forest automatically captures interactions between baseline covariates $X_i$ and treatment assignment $W_i$ through the tree  structure.\footnote{There is a caveat to this claim: Since the treatment indicator is included in the same way as the other covariates, it is  likely to be ignored in several trees that never split on it. This can cause the S-learner with Random Survival Forest to perform poorly in some situations.}  When a conventional regression is used as the predictive model, one needs to explicitly specify treatment-covariate interaction terms in order to model HTE: 

\begin{example}[S-learner with Cox-Lasso]
\label{example:lassoS}
A commonly used risk model for survival data is given by the Cox proportional hazards (PH) model with Lasso penalization~\citep{tibshirani1997lasso, goeman2010l1}. Given covariates \smash{$\widetilde{X}_i$}, the PH model assumes that the log-hazard is equal to \smash{$\beta^\intercal \widetilde{X}_i$}, for an unknown coefficient vector $\beta$ that is estimated by minimizing the negative log partial likelihood~\citep{cox1972regression} plus the sum of absolute values of the coefficient $\beta_j$ multiplied by the regularization parameter $\lambda \geq 0$ (i.e., $\lambda  \cdot \sum_j |\beta_j|$)\footnote{ 
In this note we provide some more details about fitting the Cox-Lasso: Given a survival dataset $\widetilde{\dataset}$ indexed by $\widetilde{\mathcal{I}}$ and with covariates $\widetilde{X} \in \mathbb R^{\tilde{d}}$, and a tuning parameter $\lambda \geq 0$, the Cox-PH Lasso model for predicting the survival probability at $t_0$ is fitted as follows (assuming for simplicity that there are no ties in the observed event times $U_i$)

$$
\begin{aligned}
&\widehat{\beta}_{\lambda} \in\argmin_{\beta} \left\{-\sum_{i \in \widetilde{\mathcal{I}}} \Delta_i\left[ \widetilde{X}_i^\intercal \beta -   \log\Bigg(\displaystyle \sum_{j \in \widetilde{\mathcal{I}}\,:\,U_j \geq U_i} \exp(\tilde{X}_j^\intercal \beta)\Bigg)\right] + \lambda \sum_{j=1}^{\tilde{d}} |\beta_j|\right\},\\ 
&\widehat{H}_{\lambda}(t_0) =  \sum_{i \in \widetilde{\mathcal{I}}\,:\, U_i \leq t_0} \left( \Delta_i \bigg / \sum_{j \in \widetilde{\mathcal{I}}\,:\, U_j \geq U_i} \exp(\widetilde{X}_j^\intercal \widehat{\beta}_{\lambda})\right).
\end{aligned}
$$
Then given a test covariate $\widetilde{x}$, the prediction of the survival probability at $t_0$ is given by:
$$ \exp\left\{-\widehat{H}_{\lambda}(t_0) \exp\left(\widetilde{x}^\intercal \widehat{\beta}_{\lambda}\right)\right\}.$$
In the main text (e.g., in Example~\ref{example:lassoT}), we use the following notation for the above predictive risk modeling procedure:
$$ \mathcal{M}^{\text{Cox-Lasso}}_{\lambda}\left(Y \sim \widetilde{X}; \widetilde{\dataset} \right).$$ 
}. One of the upshots of the Cox-Lasso is that it automatically performs shrinkage and variable selection;  $\lambda$ determines the sparsity of the solution (i.e., how many of the $\beta_j$ are equal to zero). In the context of CATE estimation, the Cox-Lasso is typically fitted using the covariate vector \smash{$\widetilde{X}_i = [X_i, W_i, W_i \cdot X_i]$}, that is, by explicitly including interaction terms $W_i \cdot X_i$ for the linear predictors. The S-learner with Cox-Lasso then takes the following form:
$$\widehat{\mu}(\cdot) = \mathcal{M}^{\text{Cox-Lasso}}_{\widehat{\lambda}}\left(Y \sim [X,W,W\cdot X];\; \dataset\right),\;\; \what{\tau}(x) = \what{\mu}([x, 1, 1\cdot x])-\what{\mu}([x, 0, 0\cdot x]),$$
where we make explicit in the notation that $\widehat{\lambda}$ is typically chosen in a data-driven way by minimizing the cross-validated log partial likelihood~\citep{van2006cross,simon2011regularization}.

A further challenge (beyond explicitly modeling interactions $W_i \cdot X_i$) in applying the S-learner with Cox-Lasso is that there are many possible choices with respect to normalization of covariates, interaction terms, and to applying different penalty factors to different coefficients. In our implementation we do not apply any shrinkage on the coefficient of $W_i$. We discuss the Cox-Lasso model in more detail, as well as our normalization/shrinkage choices in Supplement~\ref{subsec:suppl_lasso_implementation}. 
\end{example}

\subsection{T-learner: Risk modeling stratified by treatment}
\label{subsec:T}

Another intuitive metalearner, called the  \emph{T-learner}, fits \emph{two} risk models as a function of baseline covariates, separately for treatment and control arms.  The T-learner relies on the following characterization of the CATE function under our assumptions:
\begin{equation}
\label{eq:tlearner_identification}
\tau(x) = \mu_{(1)}(x) - \mu_{(0)}(x), \text{ where } \mu_{(w)}(x) = \mathbb P[T_i(w) > t_0 \mid X_i=x].
\end{equation}
In the notation of~\eqref{eq:risk_prediction}, the T-learner proceeds as follows:
\begin{equation}
\label{eq:tlearner}
\what{\mu}_{(1)}(\cdot) = \mathcal{M}(Y \sim X;\; \dataset_1),\;\;\;\what{\mu}_{(0)}(\cdot) = \mathcal{M}(Y \sim X;\; \dataset_0),\;\;\; \what{\tau}(x) = \what{\mu}_{(1)}(x)-\what{\mu}_{(0)}(x).
\end{equation}
In words, the model $\what{\mu}_{(1)}(\cdot)$ is fitted using data only from treated subjects, while $\what{\mu}_{(0)}(\cdot)$ is fitted using data only from control subjects. $\what{\tau}(\cdot)$ is then computed as the difference of these two models. The risk models could again be, e.g., random survival forests, or Cox-Lasso models. The following example demonstrates how the T-learner with random survival forests may be implemented using the \mintinline{R}{grf} package.

\begin{example}[T-learner with Random Survival Forest]
\label{example:grfT}
Following the code notation used in Example~\ref{example:grfS}, the T-learner takes the following form:
\begin{minted}[fontsize = \footnotesize,
               bgcolor = bg]{R}
library(grf)
m1 <- survival_forest(X[W==1,], U[W==1], D[W==1], prediction.type="Nelson-Aalen")
m0 <- survival_forest(X[W==0,], U[W==0], D[W==0], prediction.type="Nelson-Aalen")
m1_x <- predict(m1, x, t0)$predictions
m0_x <- predict(m0, x, t0)$predictions
tau_x <- m1_x - m0_x
\end{minted}
In the second line we fit the random survival forest with covariates \mintinline{R}{X} only on the treated subjects through the subsetting \mintinline{R}{W==1}. In the third line we fit the same model on control subjects (subsetting  \mintinline{R}{W==0}). Then we estimate the survival probability with covariates \mintinline{R}{x} under treatment (line 4) and control (line 5), and take the difference to estimate the CATE (line 6).
\end{example}

The next example describes the T-learner with Cox-Lasso. In contrast to the S-learner from Example~\ref{example:lassoS}, here no special tuning or normalization is required when fitting the Cox-Lasso since the risks are modeled as functions of baseline covariates only.

\begin{example}[T-learner with Cox-Lasso]
\label{example:lassoT}
The CATE estimate is computed as follows:
$$\widehat{\mu}_{(1)}(\cdot) = \mathcal{M}^{\text{Cox-Lasso}}_{\widehat{\lambda}_1}\left(Y \sim X;\; \dataset_1 \right),\;\;\widehat{\mu}_{(0)}(\cdot) = \mathcal{M}^{\text{Cox-Lasso}}_{\widehat{\lambda}_0}\left(Y \sim X;\; \dataset_0 \right),\;\; \what{\tau}(x) = \what{\mu}_{(1)}(x)-\what{\mu}_{(0)}(x).$$
Notice that the choice of penalty parameter $\widehat{\lambda}_1, \widehat{\lambda}_0$ is different for treated, resp. control groups. It may be chosen by cross-validation separately on treated subjects $\dataset_1$ and control subjects $\dataset_0$.
\end{example}

%\citet{Nie2020} addressed a similar issue named \emph{regularization bias} using an example of T-learner of lasso, where CATE may be regularized away from zero when the main effects in two models are regularized towards zero separately, giving that the true CATE is zero. In addition, \citet{Kunzel2019} also pointed out that T-learner can be extremely problematic when the number of subjects under two treatment arms are very unbalanced. Despite all the concerns mentioned above, it is easy to see that T-learner performs well under one uncommon scenario where the response functions under two treatment arms are completely unrelated. 

\subsection{Metalearning by directly modeling treatment heterogeneity}
A caveat of the S- and T-learners is that they target the statistical estimands $\mu([x,w])$ in~\eqref{eq:slearner_identification}, resp. $\mu_{(1)}(x), \mu_{(0)}(x)$ in~\eqref{eq:tlearner_identification}, but not directly the CATE $\tau(x)$ in~\eqref{eq:CATE}. As such they provide no direct way to control regularization of $\tau(x)$ and furthermore directly regularizing the risk models may lead to unintended suboptimal performance for the estimated CATE. For example, a major concern regarding the T-learner is that the two risk models $\what{\mu}_{(1)}(\cdot)$ and $\what{\mu}_{(0)}(\cdot)$ may use a different basis (predictors and their transformations)~\citep{Powers2017}, e.g., in the T-learner with the Cox-Lasso, the two models may choose a different subset of covariates. We also refer to~\citet[Figure 1]{Kunzel2019} for an iconic illustration of regularization-induced bias for the T-learner. In such cases, the difference between $\what{\mu}_{(1)}(\cdot)$ and $\what{\mu}_{(0)}(x)$ could be due to the discrepancy in model specification rather than true HTE.\footnote{The issue of regularization-induced bias becomes even more nuanced in the observational study setting of Remark~\ref{remark:observational}, see e.g., \citet{hahn2020bayesian}.}

The metalearners we describe subsequently seek to address this shortcoming of the S- and T-learners by directly modeling the CATE. Suppose---as a thought experiment---that we could observe the individual treatment effects $Y_i(1) - Y_i(0)$ for all units $i$. Then we could directly learn the CATE by predictive modeling of this \emph{oracle score}, denoted as $Y_i^{*,o}= Y_i(1) - Y_i(0)$, as a function of $X_i$ and leverage assumed properties of the CATE function (e.g., smoothness, sparsity, linearity) through the choice of predictive model and regularization strategy. 

There are two challenges in the above thought experiment: First, due to censoring we do not observe $Y_i$ for the censored $i$; instead we observe $(U_i, \Delta_i)$. Second, by the fundamental problem of causal inference, we observe at most one potential outcome (each $i$ is either treated or not), namely $Y_i = Y_i(W_i)$. It turns out, however, that there exist metalearners that can handle the ``missing data'' issue and enable direct modeling of the CATE.

\subsubsection{Censoring adjustments}
\label{subsubsec:ipcw}
As a first ingredient for the metalearners we describe below, we explain how one can account for censoring by Inverse Probability of Censoring Weights (IPCW)~\citep{kohler2002prediction, Laan2003}.

We create a dataset with only \emph{complete} cases, which are defined as subjects who had an event before $t_0$ or finished the follow-up until $t_0$ without developing an event, i.e.,  $\mathcal{I}_{\text{comp}} = \{i \in \mathcal{I} : \Delta_i=1 \text{ or } U_i \geq t_0\}$ and $\dataset_{\text{comp}} = \{(X_i, W_i, U_i, \Delta_i): i \in \mathcal{I}_{\text{comp}}\}$. \footnote{For these cases, the outcome $Y_i$ is known. To see this, note first that if $\Delta_i=1$, then we observe $T_i$ and hence we observe $Y_i = 1(T_i > t_0)$. On the other hand, if the observation is censored ($\Delta_i=0$), but $U_i \geq t_0$, then we also know that $T_i > U_i \geq t_0$, i.e., that $T_i > t_0$ and that $Y_i=1$.} Next, we estimate the survival function of the censoring time as follows:\footnote{We slightly abuse notation here since we define $S^C(u, x, w)$ as the probability of $\ind{C_i \geq u}$ given $X_i, W_i$ for all $u\geq 0$. In our implementations, however, we predict $S^C(u, x, w)$ with a strict inequality as $\mathbb{P}(C_i > u \mid X_i=x, W_i=w])$ as an approximation since the most commonly used survival models (including the ones we describe herein: Random Survival Forests, Cox-Lasso, Kaplan-Meier) follow the typical definition of a survival function, i.e., $S(t)=\mathbb{P}(T_i > t)$. We refer to~\citet{haider2020effective} for a comprehensive discussion of survival models that can predict at all $u$, and survival models that cannot.}
\begin{equation}
\label{eq:censoring_model}
\widehat{S^C}(\cdot) = \mathcal{M}^{\text{oob}}( \ind{C \geq u} \sim [X,W]; \;\dataset),\;\; \text{where } S^C(u, x, w) = \mathbb{P}[C_i \geq u \mid X_i=x, W_i=w].
\end{equation}

Based on $\widehat{S^C}(\cdot)$, we then re-weight every subject  $i \in \mathcal{I}_{\text{comp}}$ by $\what{K}_i$, an estimate of the inverse of the probability of not being censored, namely: 
\begin{equation}
\label{eq:censoring_weights}
\begin{aligned}
\what{K}_i = 1\Big/\widehat{S^C}\left(\min\{U_i, t_0\}, X_i, W_i\right).
\end{aligned}
\end{equation}
In view of~\eqref{eq:censoring_model}, the $\what{K}_i$ are estimated out-of-bag and this helps to avoid overfitting. To estimate the survival function for censoring $ S^C(u, x, w)$, we need to switch the roles of $T_i$ and $C_i$ by using the censoring indicator  $1-\Delta_i$ instead of $\Delta_i$. 

We first explain how to compute the $\what{K}_i$ with the Kaplan-Meier estimator, when censoring may be assumed to be completely independent, namely $C_i \independent (T_i,X_i, W_i)$.

\begin{example}[Censoring weights with Kaplan-Meier]
\label{example:kaplan_meier}
Let $\mathcal{F}_1,\dotsc,\mathcal{F}_L$ be a partition of $\mathcal{I}$ into $L$ folds (by default we set $L=10$). Then, for $i \in \mathcal{F}_{\ell}$, the out-of-fold Kaplan-Meier estimator of $K_i$ is equal to:\footnote{Assuming no ties for simplicity in the formula.}
$$ \what{K}_i = \left\{\widehat{S^C}_{\mathcal{F}_{-\ell}}(\min\{U_i, t_0\})\right\}^{-1} = \left\{ \prod_{\substack{k:U_k \leq \min\{U_i, t_0\} \\ k \in \mathcal{F}_{-\ell}}} \left(1\;- \;\frac{\ind{\Delta_k =0}}{|j \in \mathcal{F}_{-\ell}\,:\, U_j > U_k|} \right)\right\}^{-1},$$
where $\mathcal{F}_{-\ell} = \mathcal{U} \setminus \mathcal{F}_{\ell}$ is the set of all subjects outside fold $\mathcal{F}_{\ell}$ and $\Delta_k = \ind{\mathrm{min}(T_i, U_k)\leq C_i}$.
\end{example}

If censoring may depend on $(X_i, W_i)$, then a more complicated model is required. For example, if one is willing to assume proportional hazards, then similarly to Examples~\ref{example:lassoS} and~\ref{example:lassoT}, one could estimate $\what{K}_i$ by running the Cox-Lasso. A more nonparametric estimate is provided by random survival forests:

\begin{example}[Censoring weights with Random Survival Forests]
\label{example:grf_censoring}
Following the notation in Examples~\ref{example:grfT},\ref{example:grfS}, the \mintinline{R}{grf} package may be used as follows for IPCW.
\begin{minted}[fontsize = \footnotesize,
               bgcolor = bg]{R}
library(grf)
cen <- survival_forest(data.frame(X,W), U, 1-D, prediction.type="Nelson-Aalen")
K <- 1/predict(cen, failure.times=pmin(U,t0), prediction.times="time")$predictions
\end{minted}
In Line 2, we fit the forest with flipped event indicator \mintinline{R}{1-D} and covariates \mintinline{R}{X,W}, and in the last line we compute the vector of censoring weights  \mintinline{R}{K}. We note that \mintinline{R}{survival_forest} in the \mintinline{R}{grf} package computes out-of-bag predictions by default (\mintinline{R}{compute.oob.predictions = TRUE}).
\end{example}

If we hypothetically had access to the oracle scores $Y_i^{*,o} = Y_i(1)-Y_i(0)$ for the uncensored samples $i \in \mathcal{I}_{\text{comp}}$ and weights $\what{K}_i$ as in~\eqref{eq:censoring_model}, then we could estimate HTEs via weighted predictive modeling as $\what{\tau}(\cdot) = \mathcal{M}(Y^{*,o} \sim X; \; \dataset, \what{K})$ (see examples below). In the next subsections we describe three metalearners, the M-, X-, and R-learners that address the challenge that---even in the absence of censoring---the oracle scores $Y_i^{*,o} = Y_i(1) - Y_i(0)$ are not available, due to the fundamental challenge of causal inference.

The observation driving these methods is that for any given (observable) score $Y_i^*$ with the property
\begin{equation}
    \label{eq:ICPW_score}
    \EE{Y_i^* \mid X_i=x} = \tau(x)\; \text{  or  }\; \mathbb E[Y_i^* \mid X_i=x] \approx \tau(x),
\end{equation}
one can estimate $\widehat{\tau}(x)$ by predictive modeling of $Y_i^*$ as a function of $X_i$. The oracle score $Y_i^{*,o} = Y_i(1) - Y_i(0)$  satisfies~\eqref{eq:ICPW_score} by definition, however, it is not the only score with this property.

\begin{remark}[Doubly robust censoring adjustments]
\label{remark:doubly_robust_censoring}
The IPCW (Inverse Probability of Censoring Weights) adjustment removes censoring bias, but it can be inefficient and unstable when the censoring rate is high in a study and a majority of the censoring events happened before $t_0$. It also may be more sensitive to misspecification of the censoring model. In such cases, one can consider a doubly robust correction~\citep{tsiatis2006semiparametric} similar to the augmented inverse-propensity weighting estimator of~\citet{Robins1994}. We do not pursue such a doubly robust correction here, because it is substantially more challenging to implement with general off-the-shelf predictive models. Case-by-case constructions are possible, e.g., the Causal Survival Forest (CSF) of~\citet{cui2021estimating} uses a doubly robust censoring adjustment, and we will compare to CSF in the simulation study.
\end{remark}

\subsubsection{M-learner}
The \emph{modified outcome method (M-learner)} \citep{Signorovitch2007, Tian2014, Athey2016, Powers2017} leverages the aforementioned insight with the following score based on the \citet{horvitz1952generalization} transformation / inverse propensity weighting (IPW):
\begin{equation}
\label{eq:transformed_F}
Y^{*,M}_i = Y_i\left(\frac{W_i}{e} - \frac{1-W_i}{1-e}\right),\;\;\; \mathbb E[Y^{*,M}_i \mid X_i=x] = \tau(x).
\end{equation}
Then, the M-learner proceeds as follows:
\begin{equation}
    \label{eq:flearner}
\what{\tau}(x)= \mathcal{M}(Y^{*,M} \sim X;\; \dataset_{\text{comp}}, \what{K}),\; \text{ where } \what{K} \text{ is as in}~\eqref{eq:censoring_weights}.
\end{equation}
The predictive model for~\eqref{eq:flearner} could be any predictive model. For example, a random (regression) forest could be used~\citep{Breiman2001}.

\begin{example}[M-learner with Random Forest CATE modeling]
\label{example:grfF}
Let \mintinline{R}{K_hat} be a vector of censoring weights, derived as in Examples~\ref{example:kaplan_meier} or~\ref{example:grf_censoring}. Also let \mintinline{R}{e} be the treatment probability. Then the M-learner with random forest CATE modeling may be implemented as follows with the \mintinline{R}{regression_forest} function in the \mintinline{R}{grf} package.

\begin{minted}[fontsize = \footnotesize,
               bgcolor = bg]{R}
library(grf)
idx <- (D == 1) | (U >= t0)
Y_M <- (U > t0) * (W/e - (1-W)/(1-e))
tau_hat_forest <- regression_forest(X[idx,], Y_M[idx], sample.weights = K_hat[idx])
tau_x <- predict(tau_hat_forest, x)$predictions
\end{minted}
In Line 2 we subset to the complete cases. In Line 3 we generate the IPW response in~\eqref{eq:transformed_F}, in Line 4 we fit the random forest and finally in Line 5 we extract the estimated CATE at \mintinline{R}{x}.
\end{example} 

In the context of survival data, \citet{zhang2017mining} proposed a related M-learner approach that uses a single regression tree (rather than forest). Another alternative could be to fit the CATE with Lasso:

\begin{example}[M-learner with Lasso CATE modeling]
\label{example:feasible_lasso_cate}
If we seek to approximate $\tau(X)$ as a sparse linear function of $X_i$, we can use the Lasso~\citep{Tibshirani1996} with squared error loss:
$$(\what{\beta}_{0}, \what{\beta}_{1}) \in \argmin_{\beta_0, \beta_1} \left\{ \sum_{i \in \mathcal{I}_{\text{comp}}} \what{K}_i\cdot \left( Y_i^{*,M} - \beta_0 - \beta_1^\intercal X_i\right)^2 \, + \, \widehat{\lambda}_{\tau} \sum_{j=1}^d |\beta_{1,j}|\right\},\;\;\; \what{\tau}(x) = \what{\beta}_{0} + \what{\beta}_{1}^\intercal x,$$
where $\mathcal{I}_{\text{comp}}$ is the set of complete observations, $\what{K}_i$ is the IPC-weight in~\eqref{eq:censoring_weights}, and $Y_{i}^{*,M}$ is the M-learner (IPW) score in~\eqref{eq:transformed_F}. $\widehat{\lambda}_{\tau} \geq 0$ may be chosen by e.g., cross-validation on squared error loss. 
\end{example}

\subsection{Modeling both the risk and treatment heterogeneity}

A downside of the M-learner is the high variance (e.g., when $e \approx 0$, then the scores $Y_{i}^{*,M}$ can blow up for treated units). In this Section we describe the X- and R-learners that also directly model the CATE. By including estimated risk models in the definition of the scores, these learners can estimate the CATE with lower variance. 

\subsubsection{X-learner}
The X-learner~\citep{Kunzel2019} constructs scores that satisfy the approximate identity in~\eqref{eq:ICPW_score} that build on the following observation:
\begin{equation}
\label{eq:Xlearner_identification}
Y_i^{*,X,1} = Y_i(1) - \mu_{(0)}(X_i),\;\; \mathbb E[Y_i^{*,X,1} \mid X_i=x] = \tau(x),
\end{equation}
where $\mu_{(0)}(x) = \mathbb E[Y_i(0) \mid X_i=x]$ is defined in~\eqref{eq:tlearner_identification}. Since $\mu_{(0)}(x)$ is unknown, it needs to be estimated in a first stage (as in the T-learner~\eqref{eq:tlearner}). The X-learner thus takes the following form
\begin{equation}
\label{eq:xlearner_treatment}
\what{\mu}_{(0)}(\cdot) = \mathcal{M}(Y \sim X;\; \dataset_0),\;\;\;\what{\tau}_{(1)}(\cdot) = \mathcal{M}(Y - \what{\mu}_{(0)}(X) \sim X;\; \dataset_1\cap\dataset_{\text{comp}}, \what{K}).
\end{equation}
In words, $\what{\tau}_{(1)}$ is the CATE estimated using data from treated units, for which $Y_i(1)$ is observed, and then the unobserved $Y_i(0)$ is imputed as $\what{\mu}_{(0)}(X_i)$. The role of treatment and control groups in ~\eqref{eq:Xlearner_identification} may be switched\footnote{That is, for $Y_i^{*,X,0} =  \mu_{(1)}(X_i) - Y_i(0)$, it holds that $\mathbb E[Y_i^{*,X,0} \mid X_i=x] = \tau(x)$.} and hence analogously to~\eqref{eq:xlearner_treatment} we could consider:
\begin{equation}
\label{eq:xlearner_control}
\what{\mu}_{(1)}(\cdot) = \mathcal{M}(Y \sim X;\; \dataset_1),\;\;\;\what{\tau}_{(0)}(\cdot) = \mathcal{M}(\what{\mu}_{(1)}(X) - Y \sim X;\; \dataset_0\cap \dataset_{\text{comp}}, \what{K}).
\end{equation}

In a last stage, the X-learner combines the two CATE estimates as follows:
\begin{equation}
\label{eq:xlearner_combination}
\what{\tau}(x) = (1-e)\cdot\what{\tau}_{(1)}(x) + e\cdot\what{\tau}_{(0)}(x).
\end{equation}
The intuition here is that we should upweight~\eqref{eq:xlearner_treatment} if there are fewer treated units, and~\eqref{eq:xlearner_control} if there are more treated units.

The two CATE models in~\eqref {eq:xlearner_treatment} and~\eqref{eq:xlearner_control} may be fitted using the same methods, described e.g., for the M-learner. We provide an example using random forests (analogous to Example~\ref{example:grfF}):

\begin{example}[X-learner with Random Forest CATE Model]
\label{example:grfX}
Suppose \mintinline{R}{m1_hat}, resp. \mintinline{R}{m0_hat} are vectors of length $n$ with $i$-th entry equal to  $\what{\mu}_{(1)}(X_i)$, resp. $\what{\mu}_{(0)}(X_i)$, with the models $\what{\mu}_{(1)}(\cdot),\what{\mu}_{(0)}(\cdot)$ fitted as in the T-learner (Subsection~\ref{subsec:T}).\footnote{For example,  $\what{\mu}_{(1)}(\cdot),\what{\mu}_{(0)}(\cdot)$ could be Cox-Lasso models as in Example~\ref{example:lassoT}, or survival forests as in Example~\ref{example:grfT}. In the latter case, the following code could be used for computing \mintinline{R}{m1_hat} and \mintinline{R}{m0_hat} using \mintinline{R}{grf}: \\
\vspace{1pt}\\
\mintinline[bgcolor=bg]{R}{m1 <- survival_forest(X[W==1,], U[W==1], D[W==1], prediction.type="Nelson-Aalen")}\\
\mintinline[bgcolor=bg]{R}{m0 <- survival_forest(X[W==0,], U[W==0], D[W==0], prediction.type="Nelson-Aalen")}\\
\mintinline[bgcolor=bg]{R}{m1_hat <- predict(m1, X, t0)$predictions; m0_hat <- predict(m0, X, t0)$predictions}
} Furthermore, let \mintinline{R}{K_hat} be a vector of length $n$ corresponding to IPC-Weights estimated as in Subsection~\ref{subsubsec:ipcw}. Then the X-learner that models the CATE using the random forest function in the \mintinline{R}{grf} package may be implemented as follows. First, we fit~\eqref{eq:xlearner_treatment} by only retaining complete, treated cases (Line 2 below), constructing the estimated scores $Y_i^{*,X,1}$~\eqref{eq:Xlearner_identification} (Line 3), fitting a random forest with IPCW (Line 4) and finally extracting the model prediction at \mintinline{R}{x}:
\begin{minted}[fontsize=\footnotesize,
               bgcolor = bg]{R}
library(grf)
idx_1 <- ((D == 1) | (U >= t0)) & (W == 1)
Y_X_1 <- (U > t0) - mu0_hat
tau_hat_1 <- regression_forest(X[idx_1,], Y_X_1[idx_1], sample.weights = K_hat[idx_1])
tau_x_1 <- predict(tau_hat_1, x)$predictions
\end{minted}
We  compute \mintinline{R}{tau_x_0}~\eqref{eq:xlearner_control} analogously.\footnote{
The code would look as follows:\\ \vspace{1pt}\\
\mintinline[bgcolor=bg]{R}{idx_0 <- ((D == 1) | (U >= t0)) & (W == 0); Y_X_0 <- mu1_hat - (U > t0)}\\
\mintinline[bgcolor=bg]{R}{tau_hat_0 <- regression_forest(X[idx_0,], Y_X_0[idx_0], sample.weights = K_hat[idx_0])}
\mintinline[bgcolor=bg]{R}{tau_x_0 <- predict(tau_hat_0, x)\$predictions}}
Finally, we compute \mintinline{R}{tau_x <- (1-e)*tau_x_1 + e*tau_x_0} to combine the two estimates as in~\eqref{eq:xlearner_combination}.
\end{example} 

Other predictive models, e.g., the Lasso (as in Example~\ref{example:feasible_lasso_cate}) could also be used instead of random forests.

\subsubsection{R-learner}
The R-learner is a metalearner proposed by~\citet{Nie2020} that builds upon a characterization of the CATE in terms of residualization of $W_i$ and $Y_i$~\citep{Robinson1988}. To be concrete, we start by centering $W_i$ and $Y_i$ around their conditional expectation given $X$, that is, we consider $W_i-e$\footnote{Note that $\mathbb E[W \mid X=x] = \mathbb E[W]= \mathbb P[W=1] =e$, due to our assumption that we are in the setting of an RCT.} and $Y_i - m(X_i)$ with $m(x) = \mathbb E[Y_i \mid X_i=x]$ equal to:
\begin{equation}
\label{eq:m_marginal}
m(x) = \mathbb E[Y_i \mid X_i=x, W_i=1] \mathbb P[W_i=1] +  \mathbb E[Y_i \mid X_i=x, W_i=0] \mathbb P[W_i=0]  
 = e \mu_{(1)}(x) + (1-e)\mu_{(0)}(x).
\end{equation}
With the above definition, the following calculations hold for the expectation of $Y_i - m(X_i)$ conditionally on $X_i=x, W_i=1$:
\begin{equation}
\begin{aligned}
\EE{Y_i - m(X_i) \mid X_i=x, W_i=1} = \mu_{(1)}(x) - m(x) = (1-e)(\mu_{(1)}(x) - \mu_{(0)}(x)) = (1-e)\tau(x).
\end{aligned}
\end{equation}
Similarly $\EE{Y_i - m(X_i) \mid X_i=x, W_i=0} = -e\tau(x)$, and so, $\EE{Y_i - m(X_i) \mid X_i=x, W_i=w} = (w-e)\tau(x)$. The preceding display enables characterization of the CATE $\tau(\cdot)$ through the  loss-based representation~\citep{robins2004optimal},
\begin{equation}
\label{eq:rloss_pop}
    \tau(\cdot) \in \argmin_{\widetilde{\tau}(\cdot)} \left\{\mathbb{E}\left[\left(\{Y_i-m(X_i)\}-\{W_i-e\}\widetilde{\tau}(X_i)\right)^2\right] \right\}.
\end{equation}
\eqref{eq:rloss_pop} suggests the following CATE estimation strategy. First, we estimate the unknown $m(\cdot)$ out-of-bag~\eqref{eq:risk_prediction_oob}:\footnote{In the case without censoring, it suffices to fit a single predictive model by regressing $Y_i$ on $X_i$, that is, $\what{m}(\cdot) = \mathcal{M}^{\text{oob}}(Y \sim X;\; \dataset),$ and this is the approach suggested in~\citet{Nie2020}. However, under censoring that may depend on treatment assignment, fitting $\mathcal{M}^{\text{oob}}(Y \sim X;\; \dataset)$ becomes challenging---for example if we naively use a survival forest with covariates $X$, then the fitted model will typically be inconsistent for $m(\cdot)$. By fitting two separate predictive models to the two treatment arms, as outlined in~\eqref{eq:rlearner_init}, and taking their convex combination (weighted by the treatment probability), we overcome this challenge and may use general predictive models.
}
\begin{equation}
\label{eq:rlearner_init}
\begin{aligned}
&\what{\mu}_{(0)}(\cdot) = \mathcal{M}^{\text{oob}}(Y \sim X;\; \dataset_0),\,\what{\mu}_{(1)}(\cdot) = \mathcal{M}^{\text{oob}}(Y \sim X;\; \dataset_1),\;\;\;\what{m}(\cdot) = e \what{\mu}_{(1)}(\cdot) + (1-e)\what{\mu}_{(0)}(\cdot).
\end{aligned}
\end{equation}
The final procedure will be robust to step~\eqref{eq:rlearner_init} in the following sense: Even if $\widehat{m}(\cdot)$ is not a good approximation to $m(\cdot)$, the R-learner may perform well. In fact, if $e = 0.5$ and we estimate $\what{m}(\cdot) \approx 0$ (even though $m(\cdot) \neq 0$), the R-learner predictions are very similar to the predictions of the M-learner.\footnote{The reason is that the following characterization of $\tau(\cdot)$ in lieu of Equation~\eqref{eq:rloss_pop} also holds:
\begin{equation*}
    \tau(\cdot) \in \argmin_{\widetilde{\tau}(\cdot)} \left\{\mathbb{E}\left[\left(Y_i-\{W_i-e\}\widetilde{\tau}(X_i)\right)^2\right] \right\}.
\end{equation*}
}
On the other hand, when $\what{m}(\cdot) \approx m(\cdot)$, then centering by $\what{m}(X_i)$ helps stabilize the estimation compared to the M-learner.

We emphasize that we use out-of-bag estimates of $\what{m}(\cdot)$ in~\eqref{eq:rlearner_init}. If we fit $\what{\mu}_{(1)}(\cdot), \what{\mu}_{(0)}(\cdot)$ with \mintinline{R}{grf} survival forests, as in Example~\ref{example:grfT}, then we obtain out-of-bag estimates by default. If the Cox-Lasso is used, as in Example~\ref{example:lassoT}, then we can get out-of-bag estimates by splitting $\mathcal{I}$ into 10 folds and using out-of-fold predictions.

Second, we let $\what{K}_i$ be IPC-weights as in~\eqref{eq:censoring_weights} and  $\mathcal{I}_{\text{comp}}$ be the index set of complete cases. Finally, we estimate $\tau(\cdot)$ by fitting a  model that leads to small values of the R-learner loss $  \sum_{i \in \mathcal{I}_{\text{comp}}} \what{K}_i \left(\{Y_i-\what{m}(X_i)\}-\{W_i-e\}\what{\tau}(X_i)\right)^2.$ Such a fitting procedure is not directly accommodated by models of the form~\eqref{eq:risk_prediction}. Below, we will describe how we may achieve this task with general predictive models~\eqref{eq:risk_prediction}. However, before doing so, we provide some examples of procedures that directly operate on the R-learner loss. We start with the simplest case:

\begin{example}[R-learner for estimating a constant treatment effect]
\label{example:r_ate}
Let $\what{m}(\cdot)$ be as in~\eqref{eq:rlearner_init} and $\what{K}_i$ as in~\eqref{eq:censoring_weights}.
Suppose there is no treatment heterogeneity, that is, $\tau(x) = \text{constant}$ for all $x$.\footnote{The procedure described in this example is valid also in the presence of HTEs and asymptotically recovers the overlap-weighted average treatment effect (see e.g., \citet{crump2009dealing}). In the setting without censoring, the procedure is described also in~\citet{Chernozhukov2018}.} Then, estimating the constant $\what{\tau}$ with the R-learner loss boils down to fitting a weighted linear regression on the complete cases with response $Y_i - \what{m}(X_i)$, predictor $W_i-e$, \emph{without} an intercept, and with weights $\widehat{K}_i$, and letting $\widehat{\tau}$ be equal to the slope of $W_i-e$ in the above regression. 
\end{example}

Generalizing the above example, if we seek to approximate $\tau(\cdot)$ as a sparse linear function of $X_i$, then we may use the Lasso (compare to Example~\ref{example:feasible_lasso_cate}):

\begin{example}[R-learner with Lasso CATE modeling]
\label{example:rlasso_cate}
Let $\what{m}(\cdot)$, $\what{K}_i$ be as in Example~\ref{example:r_ate}. The R-learner estimate of the CATE using the Lasso is the following:
$$(\what{\beta}_{0}, \what{\beta}_{1}) \in \argmin_{\beta_0, \beta_1} \left\{ \sum_{i \in \mathcal{I}_{\text{comp}}} \what{K}_i\cdot \left( \left (Y_i - \what{m}(X_i)\right) - (W_i-e)\left(\beta_{0} + \beta_{1}^\intercal X_i \right)\right)^2 \, + \, \widehat{\lambda}_{\tau} \sum_{j=1}^d |\beta_{1,j}|\right\},\; \what{\tau}(x) = \what{\beta}_{0} + \what{\beta}_{1}^\intercal x.$$
The above objective can be fitted with standard software for the Lasso (e.g., \mintinline{R}{glmnet}~\citep{glmnet}) with the following specification. The response is equal to $Y_i - \widehat{m}(X_i)$, the covariates are $[W_i-e, (W_i-e)X_{i1}, \dotsc, (W_i-e)X_{id}]$, the intercept is not included, and the coefficient of the first covariate $(W_i-e)$ is unpenalized.
\end{example}

More generally, we may fit the R-learner objective using methods for varying coefficient models~\citep{hastie1993varying}. The  \mintinline{R}{causal_forest} function of the \mintinline{R}{grf} package enables direct modeling of $\tau(x)$ through the generalized random forest framework of~\citet[Section 6]{Athey2019}. For each $x$, the causal forest estimates $\what{\tau}(x)$ as in Example~\ref{example:r_ate} with additional data-adaptive weighting localized around $x$ and determined by the collection of tree splits~\citep*[Section 1.3]{athey2019estimating}.

\begin{example}[Causal Forest: R-learner with Random Forest CATE model]
\label{example:grfR}
Suppose \mintinline{R}{m1_hat}, resp. \mintinline{R}{m0_hat} are vectors of length $n$ with $i$-th entry equal to $\what{\mu}_{(1)}(X_i)$, estimated out-of-bag and let \mintinline{R}{K_hat} be a vector of length $n$ corresponding to IPC weights estimated as in Subsection~\ref{subsubsec:ipcw}. Then the R-learner that models the CATE using the random forest function in the \mintinline{R}{grf} package may be implemented as follows.
\begin{minted}[fontsize = \footnotesize,
               bgcolor = bg]{R}
library(grf)
idx <- (D == 1) | (U >= t0)
m_hat <- mu1_hat * e + mu0_hat * (1 - e)
Y <- U > t0
tau_hat <- causal_forest(X[idx,], Y[idx], W[idx], Y.hat = m_hat[idx], W.hat = e[idx], 
                         sample.weights = K_hat[idx])
tau_x <- predict(tau_hat, x)$predictions
\end{minted}
In Line 2, we get the indices of complete cases. In Line 3, we combine the estimate of $\mu_{(1)}(\cdot), \mu_{(0)}(\cdot)$ to get an estimate of $m(\cdot)$ as in~\eqref{eq:rlearner_init}. In Line 4, we compute the survival indicator \mintinline{R}{Y}. In Lines 5 and 6, we fit a causal forest that targets the R-learner objective. Note that here we subset only to complete cases given by \mintinline{R}{idx}, and we also specify the censoring weights \mintinline{R}{K_hat}, as well as the expected responses  \mintinline{R}{m_hat} for \mintinline{R}{Y}, and \mintinline{R}{e} for \mintinline{R}{W}. Finally, in Line 7, we extract the causal forest estimate of the CATE at \mintinline{R}{x}.
\end{example}

The preceding examples presented three approaches that directly operate on the R-learner loss function. It is possible, however, to cast R-learner based estimation of $\what{\tau}(\cdot)$ in the form~\eqref{eq:risk_prediction}. To do so, we rewrite~\eqref{eq:rloss_pop} equivalently as:
\begin{equation}
\label{eq:rloss_pop_equiv}
    \tau(\cdot) \in \argmin_{\widetilde{\tau}(\cdot)} \left\{\mathbb{E}\left[(W_i-e)^2\left(\frac{Y_i-m(X_i)}{W_i-e}-\widetilde{\tau}(X_i)\right)^2\right] \right\}.
\end{equation}
This is a weighted least squares objective with weights $(W_i-e)^2$. Under unbalanced treatment assignment, e.g., when there are fewer treated units ($e < 0.5$), then the R-learner upweights the treated units compared to control units by the factor $(1-e)^2/e^2$. The upweighting of treated units by the R-learner is similar to the behaviour of the X-learner (compare to~\eqref{eq:xlearner_combination}). In fact, the predictions of R-learner and X-learner (when used with the same predictive models) are almost identical in the case of strong imbalance $(e \approx 0)$~\citep{Nie2020}.

\eqref{eq:rloss_pop_equiv} justifies the following equivalent implementation of the R-learner. Let $\what{m}(\cdot)$ be as in~\eqref{eq:rlearner_init} and let $\what{K}_i$ be IPC-weights, then we may estimate $\what{\tau}(\cdot)$ as follows:
\begin{equation}
\label{eq:rlearner}
\what{Y}^{*,R} = \frac{Y - \what{m}(X)}{W - e},\;\; \what{\tau}(\cdot) = \mathcal{M}(\what{Y}^{*,R} \sim X;\; \dataset_{\text{comp}}, \what{K} \cdot (W-e)^2).
\end{equation}
We make two observations: First, $\what{Y}^{*,R}$ approximates a score in the sense of~\eqref{eq:ICPW_score}.\footnote{
Formally the score is as follows:
\begin{equation}
\label{eq:transformed_R}
Y^{*,R}_i = \frac{Y_i - m(X_i)}{W_i - e} ,\;\;\mathbb E[Y^{*,R}_i \mid X_i=x] = \tau(x).
\end{equation}
The U-learner by~\citet{Kunzel2019} is a metalearner, related to the R-learner, that directly operates on the score~\eqref{eq:transformed_R}. The first step of the U-learner is identical to the first step of the R-learner, namely $\what{m}(\cdot)$ is estimated as in~\eqref{eq:rlearner_init}. The step~\eqref{eq:rlearner}, however, is replaced by the following step:
\begin{equation*}
\label{eq:ulearner}
\what{Y}^{*,R} = \frac{Y - \what{m}(X)}{W - e},\;\; \what{\tau}(\cdot) = \mathcal{M}(\what{Y}^{*,R} \sim X;\; \dataset_{\text{comp}}, \what{K}).
\end{equation*}
The difference is in the weights; the U-learner directly uses the IPC-weights $\what{K}_i$, while the R-learner adjusts the weights  as $\what{K}_i \cdot (W_i - e)^2$, i.e., by multiplying the IPC-weights by $(W_i-e)^2$. The U-learner performed subpar compared to the R-learner in the simulation studies of~\citet{Nie2020} and so we do not consider it in our benchmark. 
}
Second, the weights used by predictive models are the product of $(W_i-e)^2$ and the IPC weights $\what{K}_i$.

To further illustrate how to apply~\eqref{eq:rlearner}, we revisit Example~\ref{example:rlasso_cate} and provide an equivalent way of implementing the R-learner with the Lasso.

\begin{example}[Weighted representation of R-learner with Lasso]
\label{example:rlasso_cate_v2} The estimator $\what{\tau}(\cdot)$ in Example~\ref{example:rlasso_cate} may be equivalently expressed as,\footnote{
Examples~\ref{example:rlasso_cate} and~\ref{example:rlasso_cate_v2} are formally equivalent as described in the main text. However, typical implementations of the Lasso first standardize all covariates to unit variance (e.g., option \mintinline{R}{standardize=TRUE} in the \mintinline{R}{glmnet} package~\citep{glmnet}). In that case, one would get conflicting answers when implementing the R-learner as described in Example~\ref{example:rlasso_cate}, respectively Example~\ref{example:rlasso_cate_v2}. In our implementation of the R-learner with the Lasso, we enable standardization and follow Example~\ref{example:rlasso_cate_v2}.
}
$$(\what{\beta}_{0}, \what{\beta}_{1}) \in \argmin_{\beta_0, \beta_1} \left\{ \sum_{i \in \mathcal{I}_{\text{comp}}} \what{K}_i\cdot (W_i-e)^2 \cdot \left( \what{Y_i}^{*,R} - \beta_{0} - \beta_{1}^\intercal X_i\right)^2 \, + \, \widehat{\lambda}_{\tau} \sum_{j=1}^d |\beta_{1,j}|\right\},\; \what{\tau}(x) = \what{\beta}_{0} + \what{\beta}_{1}^\intercal x,$$
where $\what{Y}_i^{*,R} = (Y_i - \what{m}(X_i))/(W_i - e)$. Note that this may be computed using standard Lasso software with response $\what{Y_i}^{*,R}$, covariates $X_i$, weights $\what{K}_i\cdot (W_i-e)^2$ and including an unpenalized intercept.
\end{example}

As a final example of the R-learner, we show how we may model the CATE with another popular machine learning method, namely XGBoost~\citep{chen2016xgboost} through~\eqref{eq:rlearner}.

\begin{example}[R-learner with XGBoost CATE model]
\label{example:xgboostR}
Suppose \mintinline{R}{m1_hat}, resp. \mintinline{R}{m0_hat} are vectors of length $n$ with $i$-th entry equal to $\what{\mu}_{(1)}(X_i)$, estimated out-of-bag and let \mintinline{R}{K_hat} be a vector of length $n$ corresponding to IPC-Weights estimated as in Subsection~\ref{subsubsec:ipcw}. Then the R-learner that models the CATE using the extreme gradient boosting function in the \mintinline{R}{xgboost} package may be implemented as follows.
\begin{minted}[fontsize = \footnotesize,
               bgcolor = bg]{R}
library(xgboost)
idx <- (D == 1) | (U >= t0)
m_hat <- mu1_hat * e + mu0_hat * (1 - e)
Y_R <- ((U > t0) - m_hat) / (W - e)
data <- data.frame(X = X[idx,], Y = Y_R[idx])
tau_hat <- xgboost(data = data, label = data$Y, 
                   weight = K_hat[idx] * (W[idx] - e[idx])^2, 
                   objective = "reg:linear", nrounds = 200)
tau_x <- predict(tau_hat, x)
\end{minted}
In Line 1, we load the \mintinline{R}{xgboost} package. In Line 2 we get the indices of complete cases. In Lines 3 and 4 we construct the R-learner score~\eqref{eq:transformed_R}. In Lines 5-8, we fit the XGBoost model with sample weights \mintinline{R}{K_hat[idx]*(W[idx]-e[idx])^2}. Finally, in Line 9, we extract the XGBoost estimate of the CATE at \mintinline{R}{x}.
\end{example} 

\subsection{Summary of Metalearners}

In this Section, we described 5 different metalearners, namely the S, T, X, R, and M-learners. Table~\ref{table:metalearner_summary} provides a high-level summary/overview of the estimation strategy underlying each of these learners, and what models they need to fit.

\begin{table}
\caption{High-level overview of metalearners. For each learner (S, T, X, R, and M), we describe their requirements in fitting a risk model $\what{\mu}$, censoring model $\what{S^C}$, and/or CATE model $\what{\tau}$, as well as how these predictive models are combined by each meta-learner to produce CATE estimates.}
\label{table:metalearner_summary}
\renewcommand{\arraystretch}{1.6}
\centering
\resizebox{\textwidth}{!}{\begin{NiceTabular}{llll}[colortbl-like]
 &  \cellcolor{ponyomediumyellow!30} Risk model & \cellcolor{ponyomediumblue!30}Censoring model & \cellcolor{ponyomediumpink!50}CATE model \\
\hline 
\cellcolor{gray!20}\textbf{S} & \cellcolor{ponyolightyellow!20} \begin{tabular}{l} $\what{\mu}(\cdot) = \mathcal{M}( Y \sim [X, W];\; \dataset)$ \end{tabular} &\multirow{2}{*}{\cellcolor{ponyolightblue!20} not applicable}  & \cellcolor{ponyolightpink!40} \begin{tabular}{l}$\what{\tau}(x) =  \what{\mu}([x,1]) - \what{\mu}([x,0])$\end{tabular} \\
\cmidrule(){1-2}\cmidrule(){4-4}
\cellcolor{gray!20}\textbf{T} & \multirow{2}{*}{\centering \cellcolor{ponyomediumyellow!20}  \begin{tabular}{l} \\  $\what{\mu}_{(1)}(\cdot) = \mathcal{M}( Y \sim X;\; \dataset_1)$ \\  \\  $\what{\mu}_{(0)}(\cdot) = \mathcal{M}( Y \sim X;\;\dataset_0)$ \end{tabular}}  & \cellcolor{ponyolightblue!20}  &  \cellcolor{ponyomediumpink!40}\begin{tabular}{l}$\what{\tau}(x) =  \what{\mu}_{(1)}(x) - \what{\mu}_{(0)}(x)$\end{tabular}\\
\cmidrule(){1-1}\cmidrule(){3-4}
\cellcolor{gray!20}\textbf{X} & \cellcolor{ponyomediumyellow!20} & \multirow{3}{*}{\centering \cellcolor{ponyomediumblue!20} \begin{tabular}{l} \\ 
$\what{S^C}(\cdot) =  \mathcal{M}^{\text{oob}}(\ind{C \geq u} \sim [X,W]; \dataset)$\\ \\ 
$\what{K} = 1 \big / \what{S^C}(\min\{U, t_0\}, X, W)$
\end{tabular}}  & \cellcolor{ponyolightpink!40} \begin{tabular}{l} $\what{Y}^{*,X} = (1-W)(\what{\mu}_{(1)}(X)-Y) +  W(Y-\what{\mu}_{(0)}(X))$ \\ $\what{\tau}_{(1)}(x) = \mathcal{M}(\what{Y}^{*,X} \sim X;\; \dataset_1 \cap \dataset_{\text{comp}}, \what{K})$\\
$\what{\tau}_{(0)}(x) = \mathcal{M}(\what{Y}^{*,X} \sim X;\; \dataset_0 \cap \dataset_{\text{comp}}, \what{K})$ \\
$\what{\tau}(x) = (1-e) \what{\tau}_{(1)}(x) + e\what{\tau}_{(0)}(x)$\end{tabular}\\ 
\cmidrule(){1-2}\cmidrule(){4-4}
\cellcolor{gray!20}\textbf{R} & \cellcolor{ponyolightyellow!20} \begin{tabular}{l}    $\what{\mu}_{(1)}(\cdot) = \mathcal{M}^{\text{oob}}( Y \sim X;\; \dataset_1)$ \\   $\what{\mu}_{(0)}(\cdot) = \mathcal{M}^{\text{oob}}( Y \sim X;\;\dataset_0)$ \end{tabular} &\cellcolor{ponyomediumblue!20} & \cellcolor{ponyomediumpink!40} \begin{tabular}{l}
$\what{m}(x) = e \what{\mu}_{(1)}(x) +(1-e)\what{\mu}_{(0)}(x) $\\
$\what{Y}^{*,R} =(Y - \what{m}(x))/(W-e)$ \\  $\what{\tau}(x) = \mathcal{M}(\what{Y}^{*,R} \sim X;\; \dataset_{\text{comp}}, \what{K} \cdot (W-e)^2)$\end{tabular}\\
\cmidrule(){1-2}\cmidrule(){4-4}
\cellcolor{gray!20}\textbf{M} & \multicolumn{1}{l}{\cellcolor{ponyomediumyellow!20} not applicable}  & \cellcolor{ponyomediumblue!20} & \cellcolor{ponyolightpink!40}\begin{tabular}{l}$Y^{*,M} = WY/e + (1-W)Y/(1-e)$ \\  $\what{\tau}(x)= \mathcal{M}(Y^{*,M} \sim X;\; \dataset_{\text{comp}}, \what{K})$\end{tabular}\\
\end{NiceTabular}}
\end{table}

In Section~\ref{sec:sim}, we conduct a simulation study to benchmark concrete instantiations of all these metalearners along with different choices for the underlying predictive model. We implement these approaches as how they are typically used in applied research to reflect their performance in practice. Furthermore, these concrete methods are accompanied by code~\citep{survlearners} that clarifies exact considerations required for their practical implementation.

\section{Simulation Study}\label{sec:sim}
As we saw above, different metalearners utilize different estimation strategies, namely, risk modeling, and/or direct modeling of treatment heterogeneity. Thus, certain combinations of metalearners and predictive models may be advantageous under particular data generating processes. In this section, we conduct a comprehensive simulation study to explore conditions wherein different CATE estimation methods may perform well or poorly. We evaluate a plethora of CATE estimators (Subsection \ref{sim:sim_estimators}) using the rescaled root mean squared error and Kendall's $\tau$ (Subsection \ref{sec:sim_metrics}) under multiple data generating mechanisms (Subsection \ref{sec:sim_dgps}). All results are fully reproducible with the code accompanying this article~\citep{survlearners}.

\subsection{Estimators under Comparison}
\label{sim:sim_estimators}

We compare the following estimation strategies:

\emph{Metalearners:} We implement the 5 metalearners (S, T, X, R, and M) described in Section~\ref{sec:metalearners} (summarized in Table ~\ref{table:metalearner_summary}) where we vary the risk model (survival forest, Cox-Lasso) and the CATE model (random forest, Lasso). We use 3-letter acronyms for each method, wherein the first letter corresponds to the meta-learner, the second to the risk model, and the third to the CATE model. For example, \emph{XFF} is the \underline{\emph{X}}-learner with risk models fitted by GR\underline{\emph{F}} Survival forests, and CATE fitted by GR\underline{\emph{F}} Random forests. Table~\ref{t:estimators} presents the 12 combinations of metalearners we consider, and their acronyms.  For the X,R,M metalearners that also require a censoring model, we use Kaplan-Meier IPCW by default (Example~\ref{example:kaplan_meier}). In one of the simulation settings (Subsection~\ref{setup:censor}) we also consider variants that use GRF Survival forest ICPW (Example~\ref{example:grf_censoring}) instead of Kaplan-Meier. Further details about the implementation are provided in Supplement~\ref{sec:details_implementation}, and code implementing these metalearner combinations is available in the \mintinline{R}{survlearners} package~\citep{survlearners}.

\begin{table}
\renewcommand{\arraystretch}{1.3}
\caption{Metalearner combinations considered in the simulation study. For risk and CATE models, we apply either the Cox-Lasso regression or the generalized random forest approach. For censoring models, we either employ the Kaplan-Meier estimator without covariates adjustment or the random survival forest method with variable adjustment.}
\label{t:estimators}
\centering
\begin{tabular}{lccc}
Risk \textbackslash  $\;$ CATE model & Lasso (L) & GRF (F) & None (*) \\ \hline 
Cox-Lasso (L) & XLL, RLL & ----- & SL*, TL*\\
GRF Survival Forest (F) & XFL, RFL & XFF, RFF & SF*, TF*\\
None (*) & M*L & M*F & -----
\end{tabular}
\quad \quad  \quad 
\begin{tabular}{lc}
Censoring model &  \\ \hline 
Kaplan-Meier & X, R, M \\
GRF Survival Forest & X, R, M\\
None & S, T
\end{tabular}
\end{table}

\emph{Cox-proportional hazards model (CPH):} This represents our baseline approach as it is very widely used in practice \citep{BAUM2017808, Lazar2010, Bress2021}. This method is the same as the S-learner in Example~\ref{example:lassoS} but without Lasso penalization, i.e., $\lambda=0$.

\emph{Causal Survival Forest (CSF):} This estimator, proposed by~\citet{cui2021estimating}, is similar to the RFF estimator we consider, with censoring model estimated with survival forests. The main difference, as discussed in more detail in Remark~\ref{remark:doubly_robust_censoring} is that instead of adjusting for censoring via IPCW, it implements a doubly robust adjustment. We use the CSF implementation in the \Rlogo~package \mintinline{R}{grf}.

\subsection{Performance Evaluation Metrics}
\label{sec:sim_metrics}
To assess the performance of each method, we considered two evaluation metrics:

\emph{Rescaled root mean squared error (RRMSE) $\mathbb{E}[(\what{\tau}(X)-\tau(X))^2  \mid
\what{\tau}]^{1/2}  / \mathrm{SD}(\tau(X))$:} The root mean squared error in estimating the CATE, normalized  by the standard deviation of the true CATE. The expectation is taken with respect to an independent test sample $X$ with the same distribution as the $X_i$. This metric is informative when the goal is to quantify patient-level treatment effects.

\emph{Kendall's $\tau$:} The rank correlation coefficient between predicted and true CATEs. This metric may be more suitable than the rescaled root mean squared error when it is of interest to decide how to prioritize or allocate treatment. 

The above metrics were computed in each Monte Carlo replicate of our simulations based on 5000 test samples. For each simulation setting we generated 100 Monte Carlo replicates and summarized the results through boxplots that show the $10\%, 25\%, 50\%, 75\%$, and $90\%$ percentiles of the corresponding metric.\footnote{In particular, the lower and upper hinges of the boxplots we show correspond to $10\%$, resp. $90\%$ percentiles.}

\subsection{Data Generating Processes}
\label{sec:sim_dgps}
To conduct a comprehensive comparison, our starting point was a baseline data generating process (DGP) motivated by the \emph{type 2} setup in \cite{cui2021estimating}. We then systematically varied five different characteristics of that DGP with the goal of assessing how each variation impacts different methods. Table~\ref{t:dgp} provides a high-level overview of all the DGPs we considered; in total we constructed 22 distinct DGPs for method comparison. 

The baseline DGP was as follows: We observe $n=5,000$ independent samples in the training dataset. The baseline covariates $X_i = (X_{i1},\dotsc, X_{ip})$ are independent and identically distributed with $X_{ij} \sim \mathrm{N}(0,1)$ and $p=25$. The treatment assignment was $W_i \sim \mathrm{Bernoulli}(e)$, where $e=0.5$. The survival time $T_i$ is drawn from a PH model with hazard function:
\begin{equation}
    \lambda_T(t \mid  X_i, W_i) = \exp\left( f_R(X_i) + f_{\tau}(X_i, W_i)\right) \cdot \sqrt{t}/2,\; f_R(X_i) = \beta_1{X_{i1}},\; f_{\tau}(X_i, W_i)=(-0.5-\gamma_1{X_{i2}})W_i \label{eq:simT0},
\end{equation}    
where we set the coefficients $\beta_1 = 1$ and $\gamma_1=0.5$. The censoring time $C_i$ is independent of $(T_i,W_i,X_i)$ and follows a Weibull distribution with hazard function
\begin{equation}
    \lambda_C = \kappa^\rho \label{eq:simC0},
\end{equation}
where we set $\kappa = 4$ and $\rho = 2$ for the scale and shape parameters, respectively.

\begin{table}
\centering
\caption{Data generating processes in the simulation study. DGPs were constructed by varying five characteristics of the underlying data structure. For each factor, there exists one base case and one to three variants. Risk and CATE models can be linear (Lin) or nonlinear (Nonlin) functions of $p_R$ and $p_\tau$ number of covariates, respectively. The censoring mechanism can be independent, dependent of covariates, or even unbalanced between treatment arms.}
\label{t:dgp}
\resizebox{\textwidth}{!}{\begin{tabular}{lcccc}
\hline
                 & Base Case & Variant I & Variant II & Variant III\\\hline
Model Complexity \\
\,\,Baseline risk function ($f_R$) & Lin, $p_R=1$ & Lin, $p_R=25$ & Nonlin, $p_R=1$ & Nonlin, $p_R=25$ \\
\,\,Treatment-covariate interaction ($f_\tau$) & Lin, $p_\tau=1$ & Lin, $p_\tau=25$ & Nonlin, $p_\tau=1$ & Nonlin, $p_\tau=25$ \\\hline
Treatment Heterogeneity - $\mathrm{sd}(\tau)/\mathrm{sd}(\mu_0)$\\
\,\,$f_R=\mathrm{Lin.}$, $f_\tau=\mathrm{Lin.}$ & 0.50 & 0.90 & 0.19& --\\
\,\,$f_R=\mathrm{Nonlin.}$, $f_\tau=\mathrm{Lin,}$ & 0.80 & 1.40 & 0.13& --\\
\,\,$f_R=\mathrm{Nonlin.}$, $f_\tau=\mathrm{Nonlin.}$ & 0.40 & 0.65 & 0.13& --\\\hline
Censoring \\
\,\,Censoring rate ($\rho=2$) & 30\% ($\kappa=4$) & 60\% ($\kappa=7$)& -- & --   \\
\,\,Censoring distribution ($\kappa=4$) & Regular ($\rho=2$) & Early ($\rho=1$) & -- & -- \\
\,\,Censoring dependency & $C\independent (X, W)$ & $C\sim X_1 + X_2W$ & $C\sim X_1 + X_2W +W$ & -- \\\hline 
Unbalanced Treatment Assignment ($e$) & 0.50 & 0.08 & -- & --\\\hline
\end{tabular}}
\end{table}

\subsubsection{Complexity of the Baseline Risk Function}
The model for $T_i$ considered in~\eqref{eq:simT0} is relatively simple: It is a well-specified Cox proportional hazards (PH) model with log-hazard that is linear in $(X_{i1}, W_i, W_i X_{i2})$. We evaluate how robust different CATE methods are to model misspecification and whether they can adapt to more complicated models for $T_i$. To do so, we start by increasing the complexity of the baseline risk by including a larger number of predictors in $f_R$~\eqref{eq:simT0}, by utilizing nonlinear $f_R$ such as indicator functions, or both: 
$$
\begin{aligned}
  (\text{Lin.}, p_R=25):\;\;\;\; & f_R(X_i) = \sum_{j=1}^{25} \beta_1 X_{ij} / \sqrt{p}, \;\;\;\;\;\;\; (\text{Nonlin.}, p_R=1):\;\;\;\;   f_R(X_i) = \beta_1\mathbb{1}\{X_{i1}>0.5\},\\
  (\text{Nonlin.}, p_R=25):\;\;\;\;  &f_R(X_i) = \tilde{\beta}_1 \mathbb{1}\{X_{i1}>0.5\} + \sum_{j=1}^{12} \tilde{\beta}_2\mathbb{1}\{X_{i(2j)}>0.5\}\mathbb{1}\{X_{i(2j+1)}>0.5\}.
\end{aligned}
$$
We set $\beta_1=1$ (as in the baseline DGP), and $\tilde{\beta}_1 = 0.99$, $\tilde{\beta}_2 = 0.33$. We note that the last specification $(\text{Nonlin.}, p_R=25)$ also includes second order interactions of baseline covariates in the log-hazard. Furthermore, in all cases, $X_i \in \mathbb R^{25}$, i.e., we do not change the dimension of the baseline covariates (e.g., in the case $p_R=1$, only the first feature influences the baseline risk, yet this information is not ``revealed'' to the different methods).

\subsubsection{Complexity of the CATE Function}
\label{subsubsec:complexity}
Treatment-covariate interaction terms directly determine HTEs. Hence we also increase the complexity of $f_{\tau}$ in~\eqref{eq:simT0} as follows:
$$
\begin{aligned}
  (\text{Lin.}, p_\tau=25):\;\;\;\; & f_\tau(X_i) = \sum_{j=1}^{25} (-0.5-\gamma_1 X_{ij} / \sqrt{p})W_i,\\
  (\text{Nonlin.}, p_\tau=1):\;\;\;\; & f_\tau(X_i) = (-0.5-\gamma_1\mathbb{1}\{X_{i1}>0.5\})W_i,\\
  (\text{Nonlin.}, p_\tau=25):\;\;\;\;  &f_\tau(X_i) = (-0.5-\tilde{\gamma}_1 \mathbb{1}\{X_{i1}>0.5\} - \sum_{j=1}^{12} \tilde{\gamma}_2\mathbb{1}\{X_{i(2j)}>0.5\}\mathbb{1}\{X_{i(2j+1)}>0.5\})W_i.
\end{aligned}
$$
We set $\gamma_1 = 0.5$, $\tilde{\gamma}_1 = 0.99$ and $\tilde{\gamma}_2 = 0.33$. In varying both the baseline risk through $f_R$ and the CATE complexity through $f_{\tau}$, we only consider combinations so that $f_R$ is at least as complex as $f_{\tau}$ (that when $p_{\tau}=25$, then also $p_R=25$, and when $f_{\tau}$ is nonlinear, then we also take $f_R$ to be nonlinear). This reflects the fact that the baseline risk could be arbitrarily complicated, but HTEs may be less so~\citep{Kunzel2019, Nie2020}.

\subsubsection{Magnitude of Treatment Heterogeneity}
The strength of treatment heterogeneity may also influence how easy or difficult it is to estimate CATEs. Intuitively, it may be easier for a model to detect the variation in treatment effects when the heterogeneity is strong versus weak. We simulate data under three of the DGPs above (fixing $p_R=25, p_{\tau}=1$ and varying the complexity, starting with  $f_R = \text{Lin.}, f_{\tau} = \text{Lin.}$). We then vary the parameter $\gamma_1$ in e.g.,\eqref{eq:simT0}: While the above simulations used $\gamma_1=0.5$, we also consider $\gamma_1=0$ and $\gamma_1=1$.  Table \ref{t:dgp} shows the treatment heterogeneity in each case, in which we measure it as the ratio of the standard deviation of true CATEs and the standard deviation of baseline risks. Note that the DGPs in the Variant II column correspond to zero treatment heterogeneity on the log-hazard, i.e., $\gamma_1=0$, and the heterogeneity we see is solely due to effect amplification \citep{Kent2016}. Effect amplification can be described as individuals with higher baseline risks may experience larger risk reduction on the absolute risk difference scale. Effect amplification is a major reason for the ``PATH risk modeling'' approach to CATE estimation (Subection~\ref{subsec:path} and \citet{kent2020predictive}) in which HTEs are modeled as a function of baseline risk.

\subsubsection{Censoring Mechanism}
\label{setup:censor}
An ubiquitous challenge in working with survival outcomes is the presence of right-censoring. We examine the impact of censoring mechanism in three subcases: varied censoring rates and distributions, heterogeneous censoring, and unbalanced censoring. 

We define the censoring rate as the proportion of complete cases, i.e., $|\mathcal{I}_{\text{comp}}|/N$.  We construct different censoring rates by altering the scale parameter in~\eqref{eq:simC0}. Compared to the 30\% censoring rate in the baseline DGP, roughly 60\% of the subjects are censored when $\kappa=7$. We vary the censoring distribution by setting the shape parameter $\rho =1$, which induces a slightly higher censoring rate (40\%), but more importantly, a higher proportion of subjects being censored at early times of the follow-up than in the baseline DGP.  

% Complexity of the censoring mechanism
In the baseline case and its variants above, we generate censoring times independently of baseline covariates. We also consider \emph{heterogeneous} (yet noninformative) censoring where we allow the scale parameter in the censoring time model to be a function of $(X_i, W_i)$, namely
\begin{equation*}
    \lambda_{C_i} = \kappa_i^\rho,\;\; \kappa_i = \mathrm{exp}(0.5 + \alpha X_{i1} + \delta X_{i2} W_i),
\end{equation*}
where $\alpha = 2$ and $\delta = 2$. Under this setting, subjects' censorship depends on their characteristics and treatment type.

A more interesting scenario that builds on top of heterogeneous censoring is \emph{unbalanced censoring}, that is, subjects in treated or untreated arms may be more likely to get censored. While $\kappa_i$ already depends on $W$ in the above setting, we make the censoring more  unbalanced by also including a main effect of $W_i$. 
\begin{equation*}
  \lambda_{C_i} = \kappa_i^\rho,\;\;  \kappa_i = \mathrm{exp}(1 + \alpha X_{i1} + \alpha W_i + \delta X_{2i} W_i).
\end{equation*}
Under this formulation, the censoring rate in the untreated arm is much higher than that in the treated arm (60\% vs. 30\%), which may happen e.g., if patients realize they are on the inactive treatment and drop early.

\subsubsection{Unbalanced Treatment Assignment}
Unbalanced treatment assignment may occur e.g., when drugs are too expensive to be assigned to half of the cohort. A consequence of unbalanced treatment assignment is that there may be insufficient data under one treatment arm for accurate CATE estimation. The X-learner has been shown to be more reliable than T-learner under unbalanced treatment assignment \citep{Kunzel2019}. To explore how robust all methods are to unbalanced treatment assignment, we create a scenario where only 8\% of the patients are treated, i.e., $W_i \sim \mathrm{Bernoulli}(e)$ with $e=0.08$.

\section{Simulation Results}\label{sec:sim_results}

\subsection{Description of Simulation Results}

\subsubsection{Results under Varying Baseline Risk and CATE Complexity}
\begin{figure}
\centering
\includegraphics[width=0.9\textwidth]{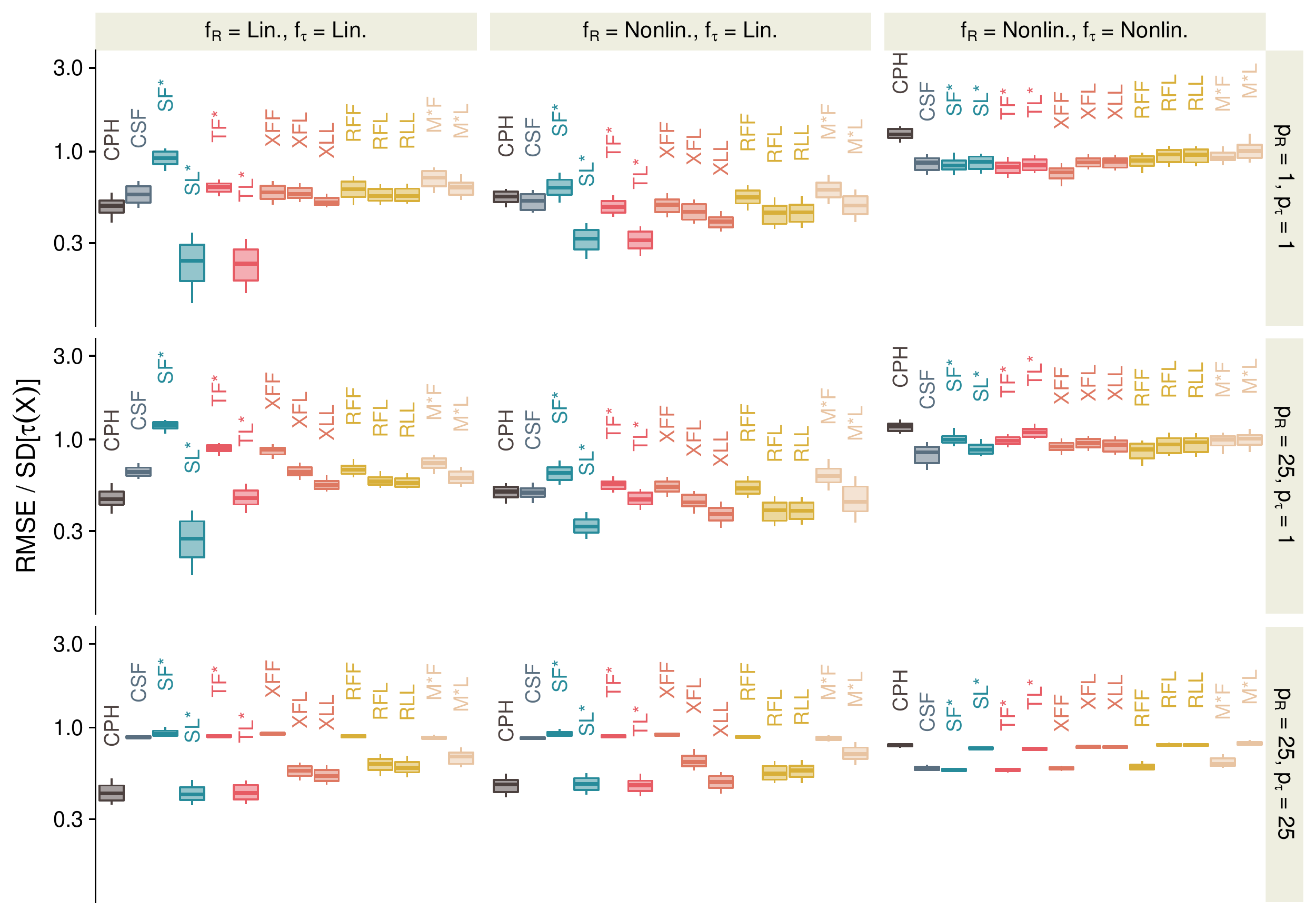}
\caption{Rescaled root mean squared errors of metalearners across various levels of complexity of baseline risk functions ($R$) and treatment-covariate interactions ($\tau$). The function forms ($f_R$ and $f_\tau$) vary between linear (Lin) and nonlinear (Nonlin) across columns, and the numbers of predictors ($p_R$ and $p_\tau$) vary across rows. We use 3-letter acronyms for each method, wherein the first letter corresponds to the meta-learner, the second to the risk model, and the third to the CATE model. For example, \emph{XFF} is the \underline{\emph{X}}-learner with risk models fitted by GR\underline{\emph{F}} Survival forests, and CATE fitted by GR\underline{\emph{F}} Random forests.}
\label{fig:complx}
\end{figure}

We first discuss the performance of the 14 CATE estimation methods across different complexity of baseline risk functions and treatment-covariate interaction terms. The rows and columns in Figure~\ref{fig:complx} vary by the numbers of covariates used in main effects ($p_R$) and interaction terms ($p_\tau$), and their functional forms, either linear (\emph{Lin}) or nonlinear (\emph{Nonlin}), respectively. 

We first make the following observations regarding the linear-linear case ($f_R=\text{Lin.}, f_{\tau}=\text{Lin.}$, first column of Figure~\ref{fig:complx}). Here, SL* and TL* outperform all other methods. The reason is that the Cox-Proportional hazards model is well-specified in this case. When either $p_R=1$ or $p_L=1$, SL* and TL* outperform CPH, i.e., the Lasso penalization helps. When all covariates are relevant ($p_R=25, p_{\tau}=25$), then CPH, SL* and TL* perform similarly. When $p_R=25, p_{\tau}=1$, SL* outperforms both TL* and CPH, i.e., it is beneficial to directly regularize the interaction effects. Finally, RLL and XLL perform quite well in absolute terms, although not as well as SL*, TL*. The reason is that RLL and XLL are misspecified: They model the CATE as linear when it is linear in terms of log-hazard ratios, but not in terms of survival probabilities.

When the baseline risk is nonlinear and the interaction term is linear ($f_R = \text{Nonlin.}, f_{\tau} = \text{Lin}$), SL* performs the best, along with XLL, RFL, RLL. When both the baseline risk and the interaction function are nonlinear ($f_R = \text{Nonlin.}, f_{\tau} = \text{Nonlin}),$ the best performing methods are the ones that model risk/CATE using GRF (CSF, SF*, TF*, XFF, RFF, M*F). Method performance in terms of Kendall's $\tau$ correlation shows a similar pattern as the RRMSEs under the same settings (Supplemental Figure \ref{fig:main_corr}). 

More broadly, this set of simulation suggests that it may be beneficial to use metalearners employing modern predictive models compared to CPH as soon as the true DGP is somewhat more complicated or has more structure compared to a dense, well-specified Cox-Proportional Hazard model. The R-learner always performs at least as well as the M-learner with the same choice of predictive model for the CATE (i.e., RFF outperforms M*F, and RLL, RFL outperform M*L). The R-learner appears to be relatively robust to the choice of model for the risk, i.e., RFL and RLL performed similarly across the settings considered (with a minor edge of RLL over RFL in the case where $f_R=\text{Lin.}, f_{\tau}=\text{Lin.})$. The choice of CATE model matters more. The X-learner appears to be more sensitive to the specification of the risk model, and the performance of XFL and XLL sometimes deviates from each other.

\subsubsection{Results under Varying HTE Magnitude}

\begin{figure}
\centering
\includegraphics[width=0.9\textwidth]{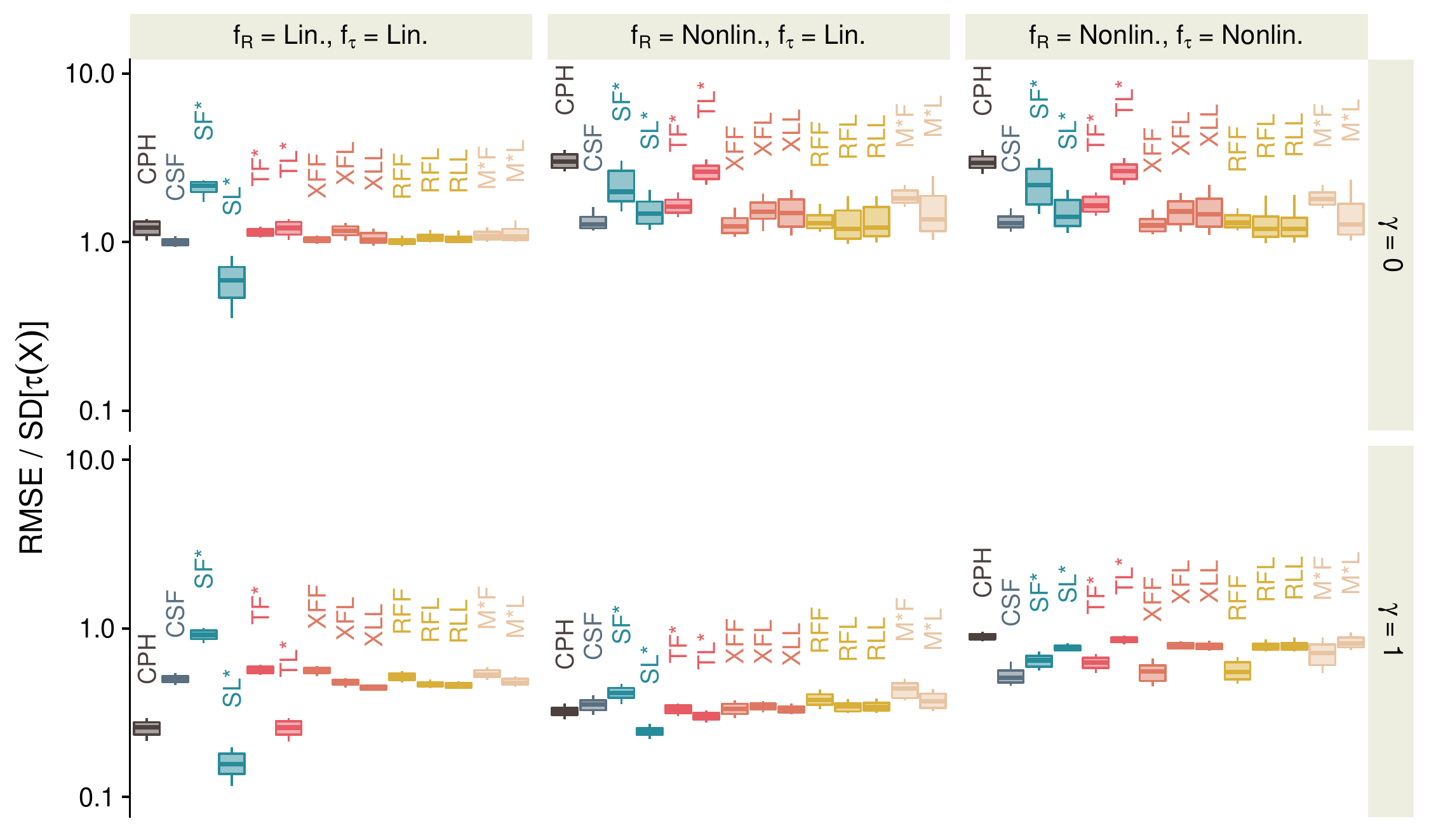}
\caption{Rescaled root mean squared errors of metalearners under various levels of treatment heterogeneity. $\gamma=0$ corresponds to zero treatment heterogeneity on the relative scale, and $\gamma=1$ yields larger heterogeneity than in the base case (Table \ref{t:dgp}). The function forms vary across three combinations of linear and nonlinear. $p_R=25$ and $p_\tau=1$.  Censoring is modeled using the Kaplan-Meier estimator. The metalearners are labeled in the same way as in Figure \ref{fig:complx}.}
\label{fig:hte}
\end{figure}

Figure \ref{fig:hte} displays the performance under varying HTE magnitude across varying levels of model complexity (with $p_R=25$ and $p_\tau=1$). All estimators show a performance drop when the treatment heterogeneity on the relative scale is zero ($\gamma=0$). When the treatment heterogeneity gets larger ($\gamma=1$), most estimators yield smaller RRMSEs; more importantly, metalearners, such as RFF or XFF, that apply machine learning approaches that are misaligned with the underlying linear risk (Figure~\ref{fig:hte}, $f_R = \mathrm{Lin}$, $f_\tau = \mathrm{Lin}$) or CATE functional forms (Figure~\ref{fig:hte}, $f_R = \mathrm{Nonlin}$, $f_\tau = \mathrm{Lin}$) now perform similarly as the estimators whose predictive models match with the true functional forms. Moreover, methods show similar performance in terms of Kendall's $\tau$ correlation when $\gamma=1$ and $f_\tau$ is linear (Figure \ref{fig:hte_corr}). 

\subsubsection{Results under Varying Censoring Mechanisms}

\begin{figure}
\centering
\includegraphics[width=0.9\textwidth]{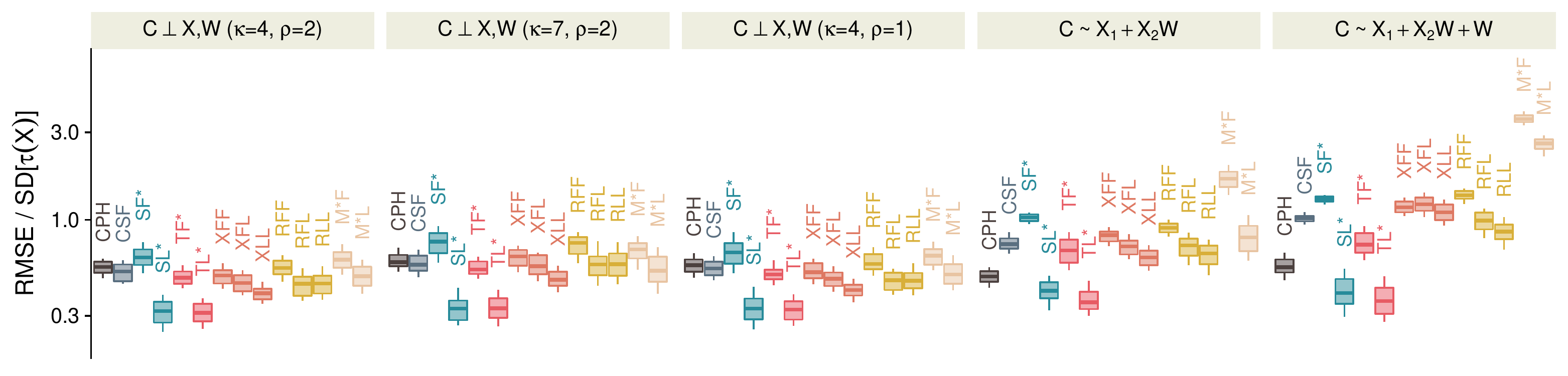}
\caption{Rescaled root mean squared errors of metalearners under varied censoring mechanism. $C\independent (X, W)$ and $C \sim (X, W)$ symbolizes censoring is \emph{not} and is a function of covariates and treatment, respectively. $\kappa$ and $\rho$ are the scale and shape parameters in the hazard function of censoring. Censoring is modeled using the Kaplan-Meier estimator (except CSF). $p_R=1$, $p_\tau=1$, $f_R = \mathrm{Nonlin}$ and $f_\tau = \mathrm{Lin}$. The metalearners are labeled in the same way as in Figure \ref{fig:complx}.}
\label{fig:cen}
\end{figure}

\begin{figure}
\centering
\includegraphics[width=0.9\textwidth]{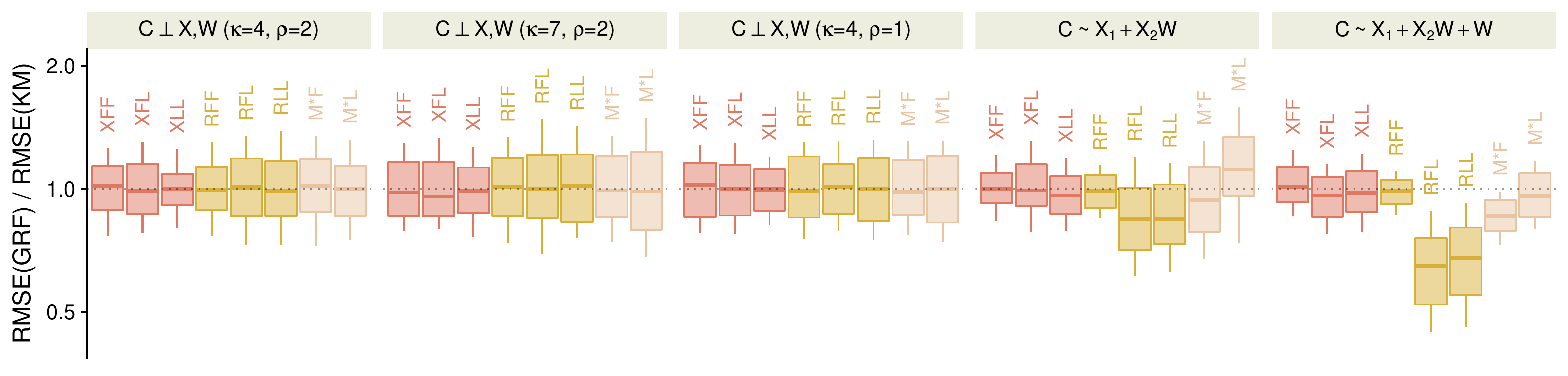}
\caption{Ratios (log-scale) of rescaled root mean squared errors of metalearners under varied censoring mechanism. The contrast is formed between using a random survival forest model versus the Kaplan-Meier estimator for modeling censoring. $C\independent (X, W)$ and $C \sim (X, W)$ symbolizes censoring is \emph{not} and is a function of covariates and treatment, respectively. $\kappa$ and $\rho$ are the scale and shape parameters in the hazard function of censoring. $p_R=1$, $p_\tau=1$, $f_R = \mathrm{Nonlin}$ and $f_\tau = \mathrm{Lin}$. The metalearners are labeled in the same way as in Figure \ref{fig:complx}.}
\label{fig:KM_SF}
\end{figure}

We then summarize method  performance under varied censoring mechanism. Figure~\ref{fig:cen} shows the RRMSE across different censoring models of all methods we have considered so far; while Figure~\ref{fig:cen_corr} shows Kendall's $\tau$. Figure~\ref{fig:censf} (resp. Figure~\ref{fig:censf_corr}) shows the RRMSE (resp. Kendall's $\tau$) of all methods when the IPC-weights are estimated by a survival forest instead of  Kaplan-Meier (this applies to the M-, X-, and R-learners). Finally, Figure~\ref{fig:KM_SF} shows the ratio of the RRMSE of each of these methods (i.e., it contrasts the performance of the estimators with survival forest ICPW compared to Kaplan-Meier IPCW).

When the censoring is independent of baseline covariates and treatment (first three panels in above figures), then both Kaplan-Meier and survival forests are well-specified. Furthermore, all methods with both IPCW adjustment perform very similarly---even though a-priori one may have anticipated Kaplan-Meier IPCW to perform better because of reduced variance). On the other hand, under heterogeneous censoring where $C \sim (X, W)$, the Kaplan-Meier estimator is misspecified. In that case, survival forest IPCW methods outperform Kaplan-Meier IPCW methods. This effect is particularly pronounced in the case of unbalanced heterogeneous censoring (panel 5). Furthermore, the X-learner appears to be more robust to misspecification of the censoring model compared to the R-learner. As we mentioned previously (also see Remark~\ref{remark:doubly_robust_censoring}), CSF and RFF with survival forest ICPW are very similar methods: The main difference is that CSF accounts for censoring with a doubly robust augmentation term, and not just IPCW (as in RFF). In our simulations, CSF and RFF perform very similarly when censoring rates are not too high; however at high censoring levels, CSF outperforms RFF.

\subsubsection{Results under Unbalanced Treatment Assignment}

\begin{figure}
\centering
\includegraphics[width=0.8\textwidth]{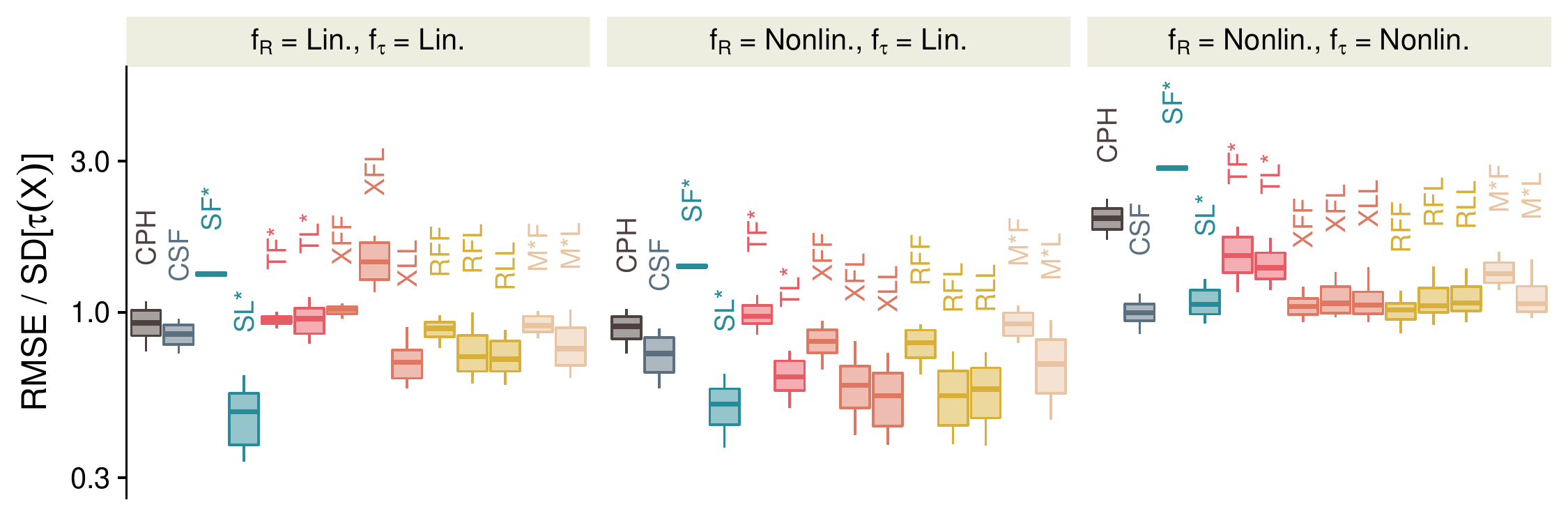}
\caption{Rescaled root mean squared errors of metalearners unbalanced treatment assignment. Only 8\% of subjects are treated. The function forms vary across three combinations of linear and nonlinear. $p_R=25$ and $p_\tau=1$. Censoring is modeled using the Kaplan-Meier estimator. The metalearners are labeled in the same way as in Figure \ref{fig:complx}.}
\label{fig:unbal}
\end{figure}

Lastly, Figures~\ref{fig:unbal} and~\ref{fig:unbal_corr} show that CATE estimation becomes more challenging under unbalanced treatment assignment. When only 8\% of the study samples are treated, all of the methods have smaller Kendalls' $\tau$ and increased median RRMSE with larger interquartile variation, and no estimator perform well when $f_R = \mathrm{Nonlin}$, $f_\tau = \mathrm{Nonlin}$. X- and R-learners, as well as CSF appear to be more robust to unbalanced treatment assignment than the other metalearners (especially when the predictive models for risk and CATE estimation are well-aligned with the true functional forms). In contrast, we observe larger performance drop in T-learners, which is consistent with earlier findings on the T-learner being sensitive to unbalanced treatment assignment~\citep{Kunzel2019}.

\subsection{Main Takeaways from Simulation Study}
\begin{figure}
\centering
\includegraphics[width=1\textwidth]{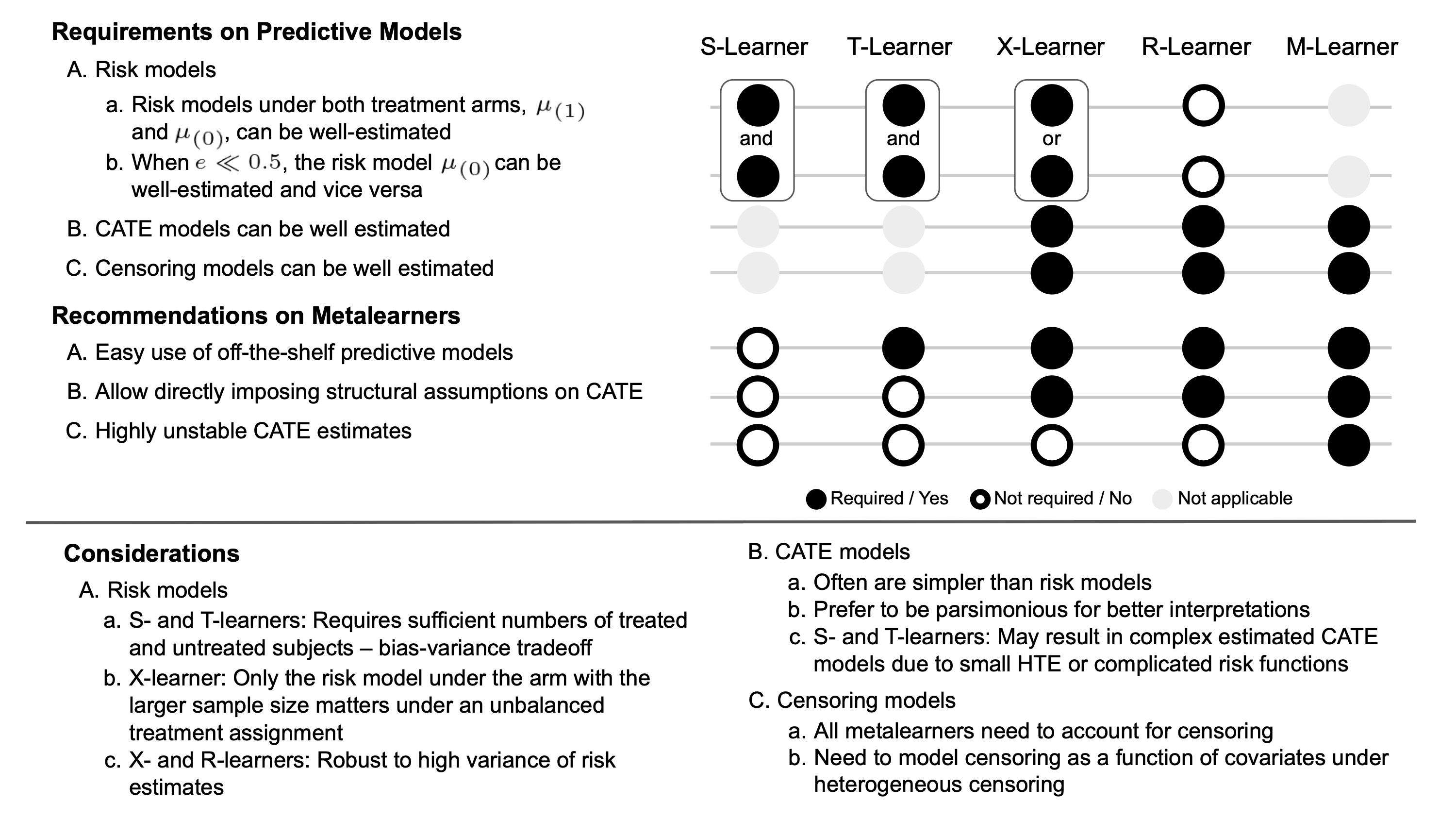}
\caption{Main takeaways from the simulation study. For each listed item on the left, we summarize metalearners by assigning three types of labels: Required/Yes, Not required/No, and Not applicable, depending on the specific requirements or recommendations described. ``\emph{and}'' indicates that both conditions need to be satisfied, and ``\emph{or}'' means that only one of the two conditions is necessary.}
\label{fig:takeaways}
\end{figure}

In this section, we summarize the major takeaways from our simulation study. We present a graphical summary of these takeaways in Figure \ref{fig:takeaways}.

First, we discuss the requirements for the risk model used with the different metalearners. The S- and T-learners require the risk models under both treatment arms to be well-estimated to provide accurate CATE estimates. Meeting this requirement is not easy in general and becomes particularly challenging under situations with unbalanced treatment assignment with a small number of subjects in one of the arms. The X- and R-learners are less sensitive to unbalanced treatment assignments. In the case of unbalanced treatment assignment, the X-learner performs well as long as the risk model for the arm with most subjects (typically, the control arm) is well-estimated. The R-learner only depends on the two risk models in order to decrease variance, and so is even more robust than the X-learner.

Second, we discuss the requirements for the CATE model. Estimating CATE functions is a crucial step for metalearners that directly model treatment effect heterogeneity, including M-, X-, and R-learners. The general intuition is that CATE functions are often simpler than risk functions, and we recommend applying parsimonious models to ensure the interpretability of CATE estimates. All three metalearners can flexibly estimate CATE functions by fitting a separate model, but M-learners tend to be unstable compared to X- and R-learners.

Third, we discuss the censoring model requirement. All approaches need to account for censoring one way or the other: S- and T-learners need to account for censoring in the process of fitting the risk models, while the other metalearners require explicit models for the censoring weights. When censoring functions are independent of baseline covariates and treatment, the Kaplan-Meier estimator is appropriate to use. However, the Kaplan-Meier estimator can induce performance degradation, when the censoring depends on baseline covariates or treatment. If it is unclear whether censoring is completely independent of treatment and covariates, we suggest to use random survival forests to model censoring as a function of relevant predictors---this choice appears to perform well even when the simpler Kaplan-Meier censoring model is correctly specified. When censoring rates are high we recommend applying the CSF~\citep{cui2021estimating} method instead of RFF.

We note that S-learners may perform well, however typically not when applied with off-the-shelf predictive models, but only with tailored models~\citep{imai2013estimating,Powers2017,hahn2020bayesian}. For instance, when used with flexible machine learning approaches such as random forests, S-learners do not give the treatment variable a special role (and so by default, some trees will not split on treatment assignment, even if HTEs are strong). When used with regression, analysts have to specify interaction terms, which typically requires substantial domain expertise~\citep{kent2020predictive}. 

To conclude, we recommend applying R- and X-learners for CATE estimation as strong default choices as any off-the-shelf machine learning models can be used. Besides, they allow imposing separate structural assumptions in the CATE and risk functions (stratified by treatment assignment), which is an important feature when these two functions are of different complexity. When background information on the control arm risk is available, analysts can apply X-learners with carefully chosen machine learning approaches that match with the possible underlying functional form. But if little is known about suitable risk models , we recommend implementing the R-learner with risk models estimated with survival forests (e.g., RFL or CSF) as default approaches as they are robust to misspecified risk models and provide stable CATE estimation.

\section{Case Study on SPRINT and ACCORD-BP}\label{sec:sprint}
The Systolic Blood Pressure Intervention Trial (SPRINT) is a multicenter RCT that evaluated the effectiveness of an intensive blood pressure (BP) treatment goal (systolic BP $<$ 120 mm Hg) as compared to the standard BP goal ($<$ 140 mm Hg) on reducing risks of cardiovascular disease (CVD). SPRINT recruited 9,361 participants and found that the intensive treatment reduced the risk of fatal and nonfatal CVD and all-cause mortality for patients at high risk of CVD events \citep{Wright2015}.

Several prior works have sought to estimate HTEs in SPRINT to improve BP control recommendations at a personalized level. For example, other authors have applied traditional subgroup analyses with either pre-specified subgroups~\citep{Cheung2812,rostomian2020heterogeneity} or subgroups learned by machine learning methods~\citep{scarpa2019assessment,wu2022heterogeneity}. More closely related to our chapter, some authors have applied variants of metalearners for HTE estimation. In particular,  \citet{patel2017personalizing} applied a S-learner with survival probabilities estimated by logistic regression (ignoring the censoring), \citet{basu2017benefit} and \citet{Bress2021} applied a S-learner along with the Cox elastic net (similar to Example~\ref{example:lassoS}, but with the elastic net~\citep{zou2005regularization} penalty instead of the Lasso penalty. \citet{Powers2017} applied S-learners along with boosting and multivariate adaptive regression splines (MARS), modified to account for the treatment $W_i$ in a specialized way. \citet{duan2019clinical} developed a X-learner with random forests for both the Risk model and the CATE model and IPC-weights estimated by a Cox-PH model.

The prior works largely suggest the existence of heterogeneous effects of intensive BP treatment, along with diverging conclusions on potential treatment effect modifiers. However, in a unified analysis using the rank-weighted average treatment effect (RATE) method, \citet{YadlowskyFlShBrWa21} found no strong evidence of existing CATE estimators being able to learn generalizable HTEs in the SPRINT and ACCORD-BP trials. The RATE can be used to evaluate the ability of any treatment prioritization rule to select patients who benefit more from the treatment than the average patient; when used to assess CATE estimators, the RATE evaluates the rule that prioritizes patients with the highest estimated CATE.

We build on this analysis by considering the behavior of a number of metalearners on these trials. We first examine the performance of metalearners in estimating CATEs in SPRINT and the ACCORD-BP trial by conducting a  global null analysis where, by construction, the CATE is zero for all subjects. Then, we analyze the CATEs in the SPRINT and ACCORD-BP data building upon insights leaned from our simulation study. Overall, our findings mirror those of \citet{YadlowskyFlShBrWa21}, i.e., we do not find significant evidence of the ability of CATE estimators to detect treatment heterogeneity in these trials.

In all our analyses we seek to estimate the CATEs, defined as the difference in survival probabilities at the median follow-up time (i.e., 3.26 years). Our outcome is a composite of CVD events and deaths, which includes nonfatal myocardial infarction (MI), acute coronary syndrome not resulting in MI, nonfatal stroke, acute decompensated congestive heart failure, or CVD-related death. We identified 13 predictors from reviewing prior works \citep{goff2014, Wright2015, basu2017benefit}, including age, sex, race black, systolic BP, diastolic BP, prior subclinical CVD, subclinical chronic kidney disease (CKD), number of antihypertensives, serum creatinine level, total cholesterol, high-density lipoprotein, triglycerides, and current smoking status. After retaining only subjects with no missing data on any covariate, the SPRINT sample includes 9,206 participants (98.3\% of the original cohort) 

In addition to the main predictor set (13 predictors) used above, we also consider a second, reduced predictor set with only 2 predictors. The first predictor is the 10-year probability of developing a major ASCVD event (``ASCVD risk'') predicted from Pooled Cohort Equations (PCE)~\citep{goff2014}, which may be computed as a function of a subset of the aforementioned 13 predictors. Including the PCE score as a predictor is justified by the ``PATH risk modeling'' framework (Subsection~\ref{subsec:path} and~\citet{kent2020predictive}) according to which absolute treatment effects are expected to be larger for larger values of the PCE score. Such evidence was provided for the SPRINT trial by~\citet{phillips2018impact} who conducted subgroup analyses with subgroups stratified by quartiles of PCE scores. The second predictor is one of the 13 original predictors, namely the binary indicator of subclinical CKD. A subgroup analysis by CKD was pre-specified in the SPRINT RCT and considered, e.g., in~\citet{Cheung2812,rostomian2020heterogeneity}.

Beyond the reanalysis of SPRINT, we also apply some of our analyses to the ACCORD-BP trial~\citep{Cushman2010} that was conducted at 77 clinical cites across U.S. and Canada. ACCORD-BP also evaluated the effectiveness of intensive BP control as in SPRINT, but one major difference is that all subjects in ACCORD-BP are under type 2 diabetes mellitus. Moreover, ACCORD participants are slightly younger (mean age = 62.2 years) than SPRINT subjects (mean age = 67.9 years) and are followed for a longer time on average (mean follow-up time = 3.3 years for SPRINT vs. 4.7 years for ACCORD-BP). Our study sample contains 4,711 patient-level data (99.5\% of the original data).

\subsection{Global Null Analysis} 
A large challenge in evaluating CATE methods using real data is the lack of ground truth. One situation in which the true CATE is known is under the global null, i.e., when all treatment effects are equal to $0$. We implement the global null analysis by restricting our attention only to subjects in the control arm of SPRINT, and assign them an artificial and random treatment $Z \sim \mathrm{Bernoulli}(0.5)$. Hence by construction $Z$ is independent of $(T_i(1),T_i(0), X_i, C_i, W_i)$ and the CATEs must be null. We then estimate the CATE with respect to the new treatment $Z$ using all the CATE approaches compared in Section \ref{sec:sim}.  We repeat the same global null analysis by using all treated subjects in SPRINT, and for the treated and untreated participants in ACCORD-BP, separately. 

In each global null analysis, we model censoring using two approaches: The Kaplan-Meier estimator or the random survival forest method via \mintinline{R}{grf}. Table \ref{t:zerocatefull} shows that some metalearners yielded much larger RMSEs than the others, which may indicate they are more likely to detect spurious HTE when it does not exist. The S-learner with GRF (SF*), M-learner with Lasso (M*L), and R-learners show consistently small RMSEs. The X-learner with GRF (XFF), which is methodologically similar to the method applied by \cite{duan2019clinical} for HTE estimation in SPRINT, result in nontrivial RMSEs. When a survival forest model is used to estimate the censoring weights versus the Kaplan-Meier estimator, the RMSEs of most metalearners decrease, which may be an indication that the censoring mechanism in SPRINT depends on baseline covariates. Table \ref{t:zerocatepce} (Supplement \ref{appendix:NULL}) shows similar results when the PCE scores and subclinical CKD are used as predictors.

Figure~\ref{fig:sprint_NULL_rmse} compares the performance of different methods under the two considered choices of predictor sets. For the global null analysis conducted with untreated units on both SPRINT and ACCORD-BP, we make the following observation: The learners that applied a Lasso model for the CATE function have a decreased RMSE when the reduced predictor set is used. This observation is concordant with the ``PATH risk modeling'' recommendations~\citep{kent2020predictive}. In contrast, the learners that used GRF for the CATE function have a lower RMSE when the main predictor covariate set is used. A possible explanation is that the presence of multiple uninformative predictors drives different trees of the forest to split on different variables; upon averaging across trees the estimated CATEs are approximately zero (as they should be under the global null). 

\begin{figure}[ht!]
\centering
\includegraphics[width=0.9\textwidth]{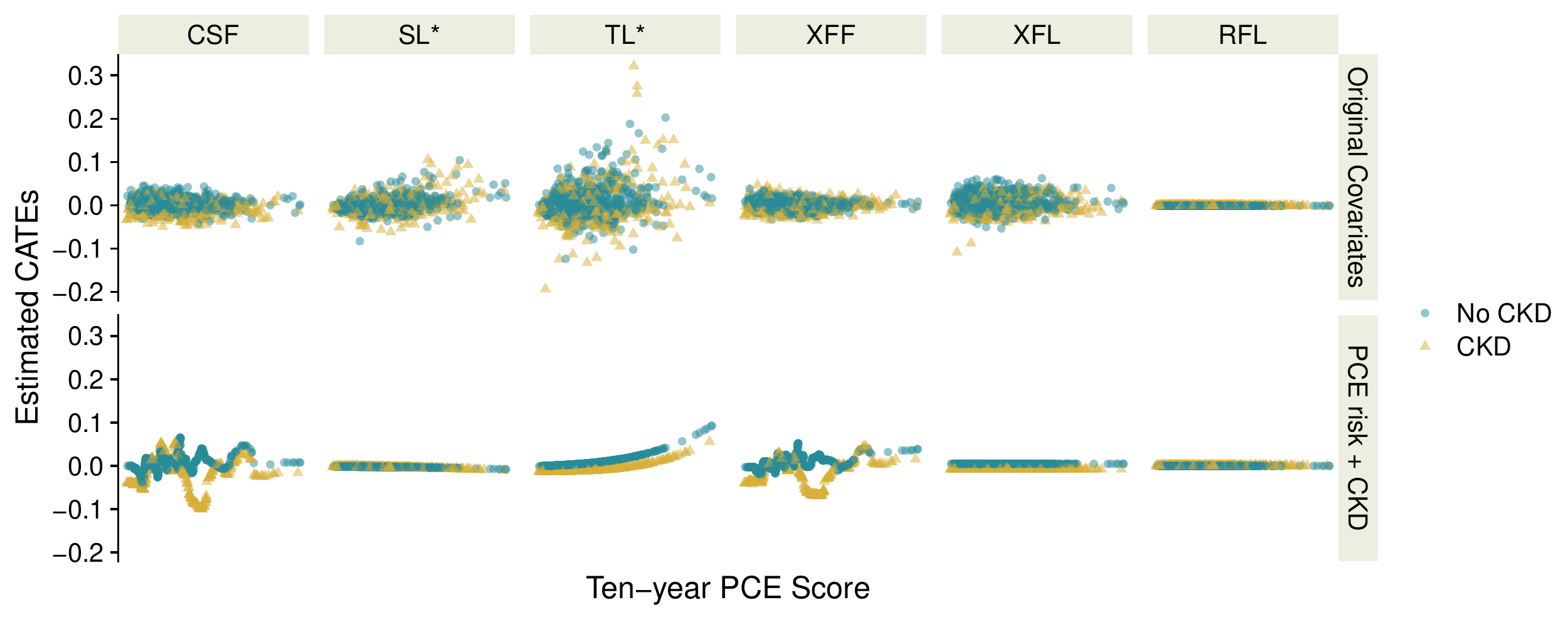}
\caption{Scatter plot of ten-year PCE scores and estimated CATEs in SPRINT under a global null model. The CATE models were derived using 70\% of the untreated patients with an artificial randomized treatment assignment then used to make predictions on the rest 30\% test data. The censoring weights are estimated using a survival forest model. The analysis was replicated with the original covariates in SPRINT as the predictors (Row 1) and the estimated ten-year CVD risk (using pooled cohort equation) and subclinical CKD as the predictors (Row 2).}\vspace{-1em}
\label{fig:sprint_NULL}
\end{figure}

Figure \ref{fig:sprint_NULL} displays the relationship between the estimated CATEs in untreated SPRINT subjects and the ten-year PCE scores for a subset of the methods that use survival forest IPCW. We see that RFL estimate constants CATEs (that is, all CATEs are equal to the estimated ATE) that are nearly zero. Under the global null, this behavior is desirable to avoid detection of spurious HTEs and showcases the benefit of R-learners in enabling direct modeling of CATEs and imposing, e.g., sparsity assumptions. Other approaches yielded non-zero CATEs with large variations. In particular, when the PCE scores and CKD are used as predictors, T-learner with Lasso (TL*) shows an increasing trend as a function of PCE scores for both subjects with or without prior subclinical CKD. This illustrates the regularization bias of the T-learner, as pointed out by \cite{Nie2020}, and explains the large RMSE of TL*. Figures~\ref{fig:sprint_NULL_ctl_KM}--\ref{fig:accord_NULL_trt_KM} are analogous to Figure~\ref{fig:sprint_NULL} and present the other 3 settings (treated units in SPRINT, untreated and treated units in ACCORD-BP) for both Kaplan-Meier IPCW and survival forest IPCW.

\input{Tables/Table4.tex}

\vspace{1em}

\subsection{CATE Estimation in SPRINT and ACCORD}
For our main SPRINT analysis, we applied five of the previously considered methods, namely CSF, TL*, XFF, XFL, and RFL. We computed CATEs with respect to the actual (rather than null) treatment assignment in SPRINT and estimated the censoring weights using a random survival forest model. For internal validation, we randomly sampled 70\% of the data as the training set and made predictions on the remaining testing set. For external validation, we employed the entire SPRINT data as the training set and made predictions on ACCORD-BP. 

As ground truth CATEs are no longer available in this setting, we assess method performance in real data using two recently proposed metrics. RATEs \citep{YadlowskyFlShBrWa21} constitute a method for evaluating treatment prioritization rules, under which a subset of the population is prioritized for the allocated treatment. The RATE method quantifies the additional treatment benefits gained by using a prioritization rule to assign treatment relative to random treatment assignment, measured as the area under the target operating curve (AUTOC). An AUTOC of zero for a prioritization rule indicates that using that rule would result in the same benefits as assigning treatment randomly within the population (i.e., average treatment effects), which further implies that either there is no treatment heterogeneity or the treatment prioritization rule does not effectively stratify patients in terms of treatment benefit. The RATE assessment can be considered as a discrimination metric for CATE estimates, as a prioritization rule with a large, positive AUTOC effectively distinguishes patients with greater treatment benefits from those with lesser treatment benefits by assigning them a high versus low treatment priority, respectively. The Expected Calibration Error for predictors of Treatment Heterogeneity (ECETH) \citep{Xu2022} is another novel metric for quantifying the $\ell_2$ calibration error of a CATE estimator. The calibration function of treatment effects are estimated using an AIPW (augmented inverse propensity weighted) score, which makes the metric robust to overfitting and high-dimensionality issues.

\begin{figure}[ht!]
\centering
\includegraphics[width=0.9\textwidth]{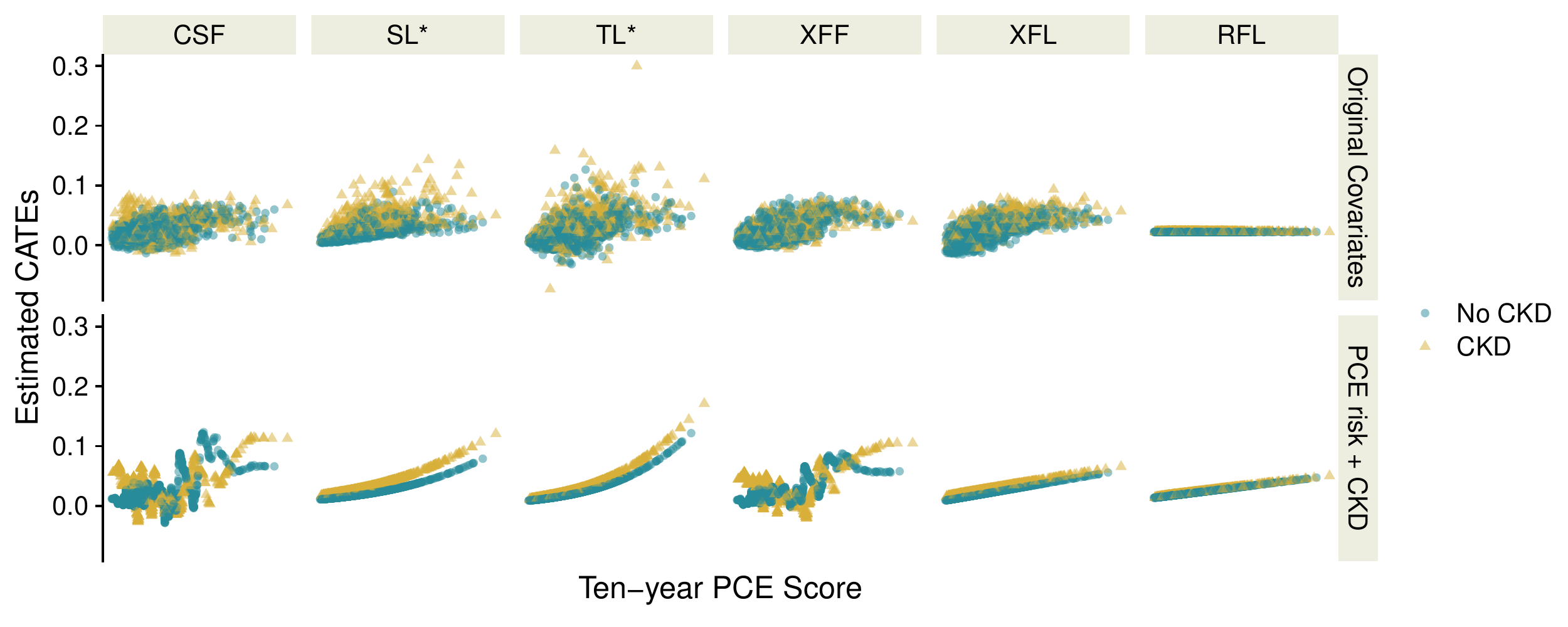}
\caption{Scatter plot of ten-year PCE score and estimated CATEs in SPRINT. This figure is analogous to Figure~\ref{fig:sprint_NULL}, but with the full SPRINT cohort and using the true treatment assignment.} \vspace{-1em}
\label{fig:sprint_full}
\end{figure}

Table \ref{t:sprint_eval} shows that none of the methods enable statistically significant detection of treatment heterogeneity in SPRINT or ACCORD-BP; in other words, RATE evaluation does not provide evidence for treatment heterogeneity in this setting (a handful of methods in Table \ref{t:sprint_eval} would be marginally significant on their own, but together the results would not pass a standard multiple testing correction). In Figure \ref{fig:sprint_full}, the CATEs estimated using RFL also show an independent relationship with the PCE score when the original predictor set is used. But when the PCE score used as a predictor, the CATE estimates from all methods showed an overall increasing trend with the ten-year PCE score, consistent with e.g., the finding of~\cite{phillips2018impact}. Such a trend is less obvious in the external validation results (Figure \ref{fig:accord_full}), and all the resulting AUTOCs are nonpositive (under PCE + CKD), which suggest that the prioritization rules based on estimated CATEs in ACCORD-BP lead to similar treatment benefits as average treatment effects. 

\input{Tables/Table5.tex}

\vspace{1em}

\section{Discussion}\label{sec:dis}
Given the increasing interest in personalized medicine, a number of advanced statistical methods have been developed for estimating CATEs, often referred to as personalized treatment effects. We focused on characterizing the empirical performance of metalearners in an RCT setting. An important direction for future work is to extend our tutorial and benchmark to the observational study setting where confounding plays a crucial role. In~\eqref{eq:CATE} we considered treatment effects in terms of difference in survival probabilities. CATEs can be also be defined on the relative scale such as hazard ratios and as the difference in restricted mean survival times. The latter is a popular estimand as it can be measured under any distribution of survival times and has a straightforward interpretation, that is, the expected life expectancy between an index date and a particular time horizon \citep{Lu2020}. Further investigations on metalearners targeting such estimands may improve HTE estimation in clinical settings.

The current work aims to provide guidance on \emph{when} and \emph{how} to apply each approach for time-to-event outcomes in light of the specific characteristics of a dataset. We conducted comprehensive simulation studies to compare five state-of-art metalearners coupled with two predictive modeling approaches. We designed a spectrum of data generating processes to explore several important factors of a data structure and summarized their impacts on CATE estimation, as well as the weakness and strengths of each CATE estimator. Based on our findings from the simulation study, we highlighted main takeaways as a list of requirements, recommendations, and considerations for modeling CATEs, which provides practical guidance on how to identify appropriate CATE approaches for a given setting, as well as a strategy for designing CATE analyses. Finally, we reanalyzed the SPRINT and ACCORD-BP studies to demonstrate that some prior findings on heterogeneous effects of intensive blood pressure therapy are likely to be spurious, and we present a case study to demonstrate a proper way of estimating CATEs based on our current learning. To facilitate the implementation of our recommendations for all the CATE estimation approaches that we investigated and to enable the reproduction of our results, we created the \Rlogo~package survlearners~\citep{survlearners} as an off-the-shelf tool for estimating heterogeneous treatment effects for survival outcomes. 

%Researchers may not have all the background knowledge mentioned in the model requirements and considerations list or they are not confident about existing information; in these cases, we suggest to narrow down to several CATE estimators. If researchers have very limited knowledge on data, we recommend applying the default approaches (R-learners and causal survival forest) that have been shown robust performance across various situations. 

%Moreover, even though our simulation study covered a wide range of data structures, the survival and censoring times were generated using proportional hazards models, so future efforts are needed to explore other distributions that do not satisfy the PH assumption. Also, a variety of statistical learning approaches are available for building outcome or CATE predictive models such as elastic net and convolutional neural networks; thus, one should also consider other alternatives than LASSO and GRF and pick the best fit for data. Last but not the least,
\printendnotes[custom]

\section*{Acknowledgements}
This work was supported by R01 HL144555 from the National Heart, Lung, and Blood Institute (NHLBI).

\setcounter{page}{1}
\renewcommand{\thepage}{S\arabic{page}} 
\setcounter{table}{0}
\renewcommand{\thetable}{S\arabic{table}}%
\setcounter{figure}{0}
\renewcommand{\thefigure}{S\arabic{figure}}
\setcounter{equation}{0}
\renewcommand{\theequation}{S\arabic{equation}}

\begin{appendices}
\section{Details for concrete implementation of the benchmarked metalearners}
\label{sec:details_implementation}
In this Supplement we provide some of the concrete details and considerations that pertain to our implementation of metalearners in Table~\ref{t:estimators}. For full details, we refer to the code implementing the different metalearners, which is available in the \mintinline{R}{survlearners} package~\citep{survlearners}.

\subsection{Lasso predictive models}
\label{subsec:suppl_lasso_implementation}

When we apply the Lasso approach to estimate risk or CATE functions, the implementation is via the \Rlogo~package \mintinline{R}{glmnet}~\citep{friedman2010regularization,glmnet}, and we choose the model that yields the minimum mean (10-fold) cross-validated error (i.e., \mintinline{R}{lambda.min}), which imposes a smaller regularization than when the default \mintinline{R}{lambda.1se} is used. Moreover, to obtain out-of-bag estimates of the risk in R-learner with Lasso, we create 10 random folds of the original data and fit 10 different models, each with a different \mintinline{R}{lambda.min}.

When we regularize CATE functions, we do not want to penalize the average treatment effects. To ensure this, in M-, X-, and R-learners, we can implement the Lasso with default settings, i.e., an intercept term is automatically included and not being penalized. By default \mintinline{R}{glmnet} also standardizes the design matrix with the covariates.
 
In the S-learner with the Cox-Lasso, one needs to pay special attention to the feature standardization, and to penalization. In particular, one needs to be careful about the strength of penalization applied to the baseline covariate effects, the main treatment effects, and the interaction effects~\citep{imai2013estimating}. In our implementation we make the concrete choices: We set the \mintinline{R}{penalty.factor} to 0 for the main effect of treatment and to 1 for the main effects of covariates and treatment-covariate interaction terms. Furthermore, we do not use the built-in function in \mintinline{R}{glmnet} to standardize (center to mean 0 and scale to variance 1) the design matrix of covariates as that will also standardize the treatment variable; instead, we standardize the baseline covariate matrix and create interaction terms before including them into a model. We also recode treatment levels to 0.5 and -0.5 (that is, $\widetilde{W}_i=W_i - 0.5$) to form modified covariates \citep{Tian2014}.

In standard linear regression, one common practice for conducting post-selection inference is via data splitting. Specifically, the data is randomly split into two halves, one half is used for model selection, and the other half is used for conduct inference. \citet{lee2016exact} developed a general approach to forming valid confidence intervals for the non-zero coefficients in the model selection by Lasso.

\subsection{GRF predictive models}
We employ a random survival/regression forest model for estimating risks, censoring weights, and CATEs. We use default values for all the tuning parameters in most cases expect the \mintinline{R}{prediction.type} and \mintinline{R}{alpha}. When we model risk and censoring with the GRF survival forest, we set \mintinline{R}{prediction.type = "Nelson-Aalen"}. This is particularly useful for censoring models as it ensures positive censoring weights. When we estimate risks with the GRF survival forest under situations with low event rates (e.g., 5\% in SPRINT data), we set the tuning parameter \mintinline{R}{alpha = 0.01} so it allows further splits as long as the number of events in each child node is at least 1\% of the sample size in the parent node.

\section{Additional Simulation Results and Figures}
\begin{figure}[H]
\centering
\includegraphics[width=0.9\textwidth]{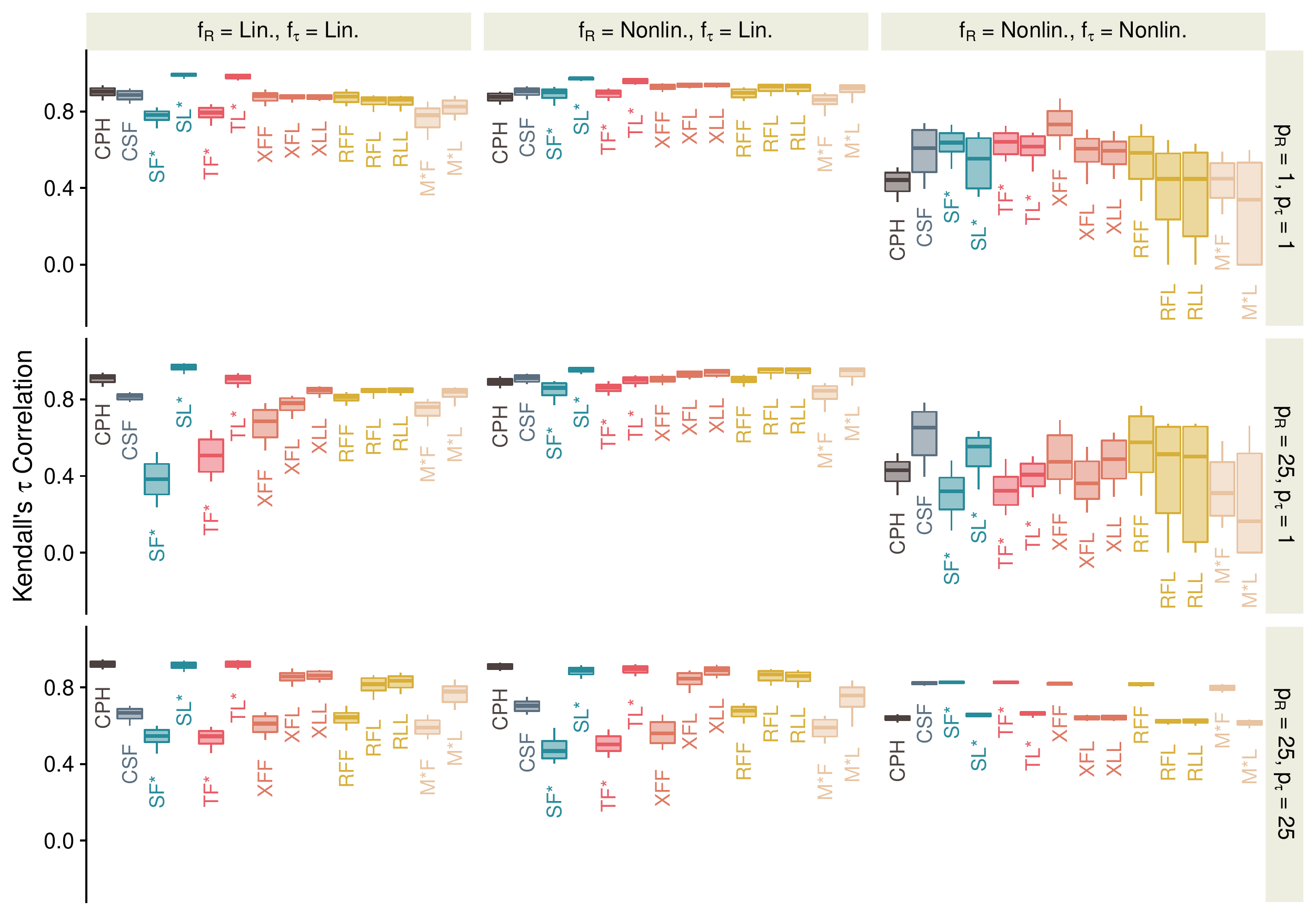}
\caption{Kendall's $\tau$ correlations of metalearners under various levels of complexity of baseline risk and CATE functions. DGPs and estimators are the same as in Figure \ref{fig:complx}.}
\label{fig:main_corr}
\end{figure}

\begin{figure}[H]
\centering
\includegraphics[width=0.9\textwidth]{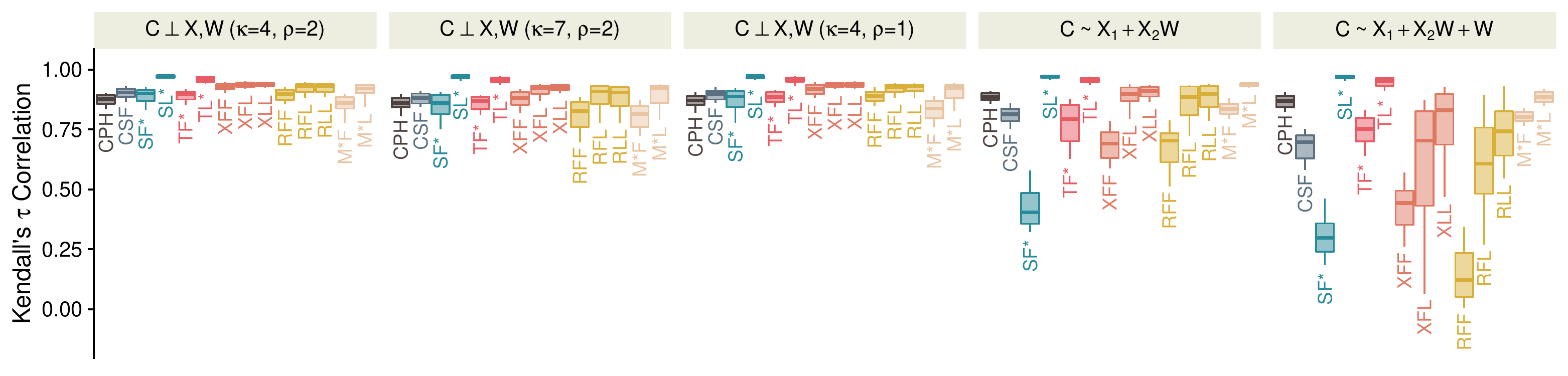}
\caption{Kendall's $\tau$ correlations of metalearners under varied censoring mechanism. DGPs and estimators are the same as in Figure \ref{fig:cen}.}
\label{fig:cen_corr}
\end{figure}

\begin{figure}[H]
\centering
\includegraphics[width=0.9\textwidth]{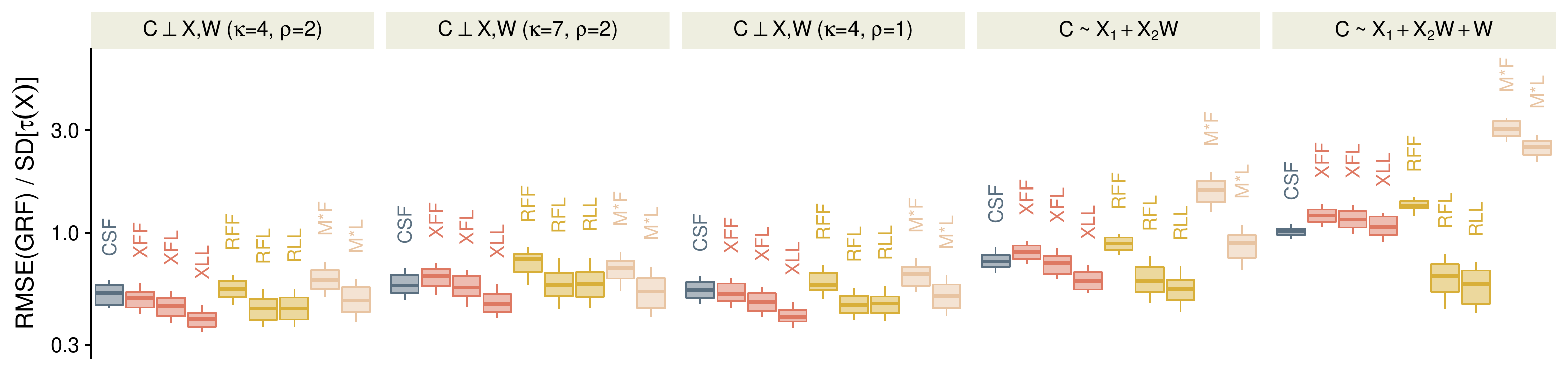}
\caption{Rescaled root mean squared errors of metalearners under varied censoring mechanism. Censoring is modeled using a random survival forest approach as compared to the results in Figure \ref{fig:cen}.}
\label{fig:censf}
\end{figure}

\begin{figure}[H]
\centering
\includegraphics[width=0.9\textwidth]{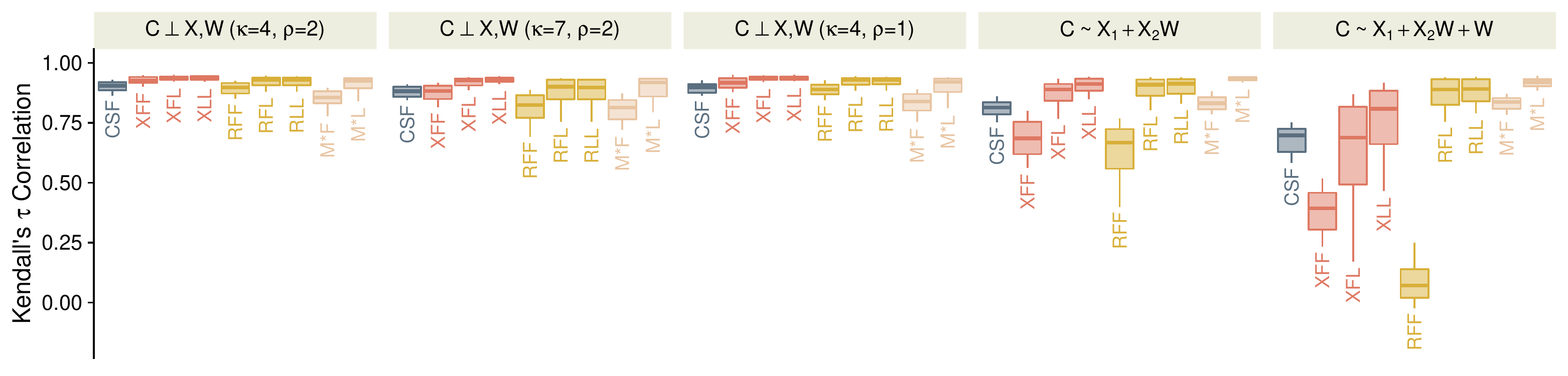}
\caption{Kendall's $\tau$ correlations of metalearners under varied censoring mechanism. DGPs and estimators are the same as in Figure \ref{fig:censf}.}
\label{fig:censf_corr}
\end{figure}

\begin{figure}[H]
\centering
\includegraphics[width=0.9\textwidth]{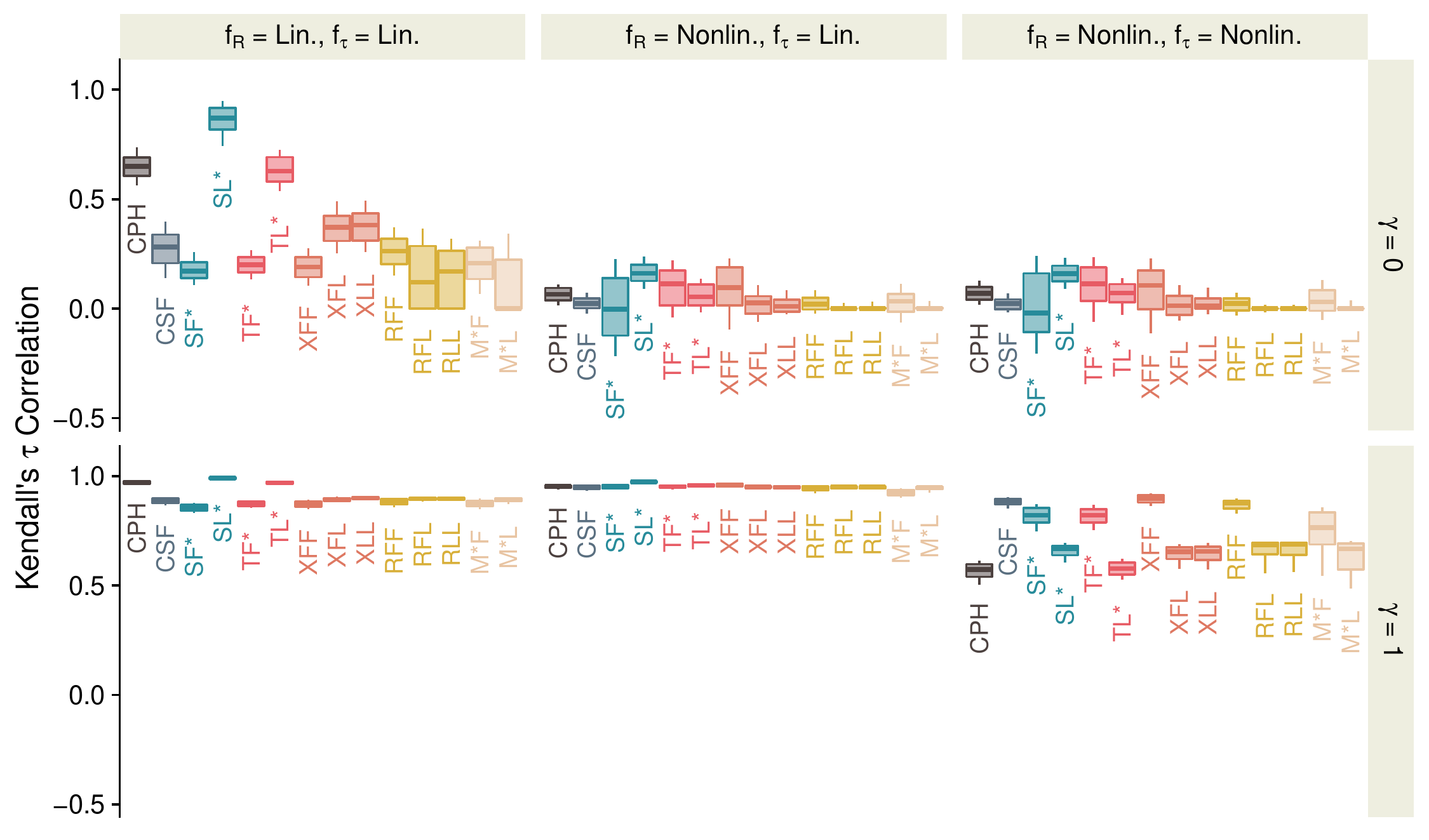}
\caption{Kendall's $\tau$ correlations of metalearners under various levels of treatment heterogeneity. DGPs and estimators are the same as in Figure \ref{fig:hte}.}
\label{fig:hte_corr}
\end{figure}

\begin{figure}[H]
\centering
\includegraphics[width=0.9\textwidth]{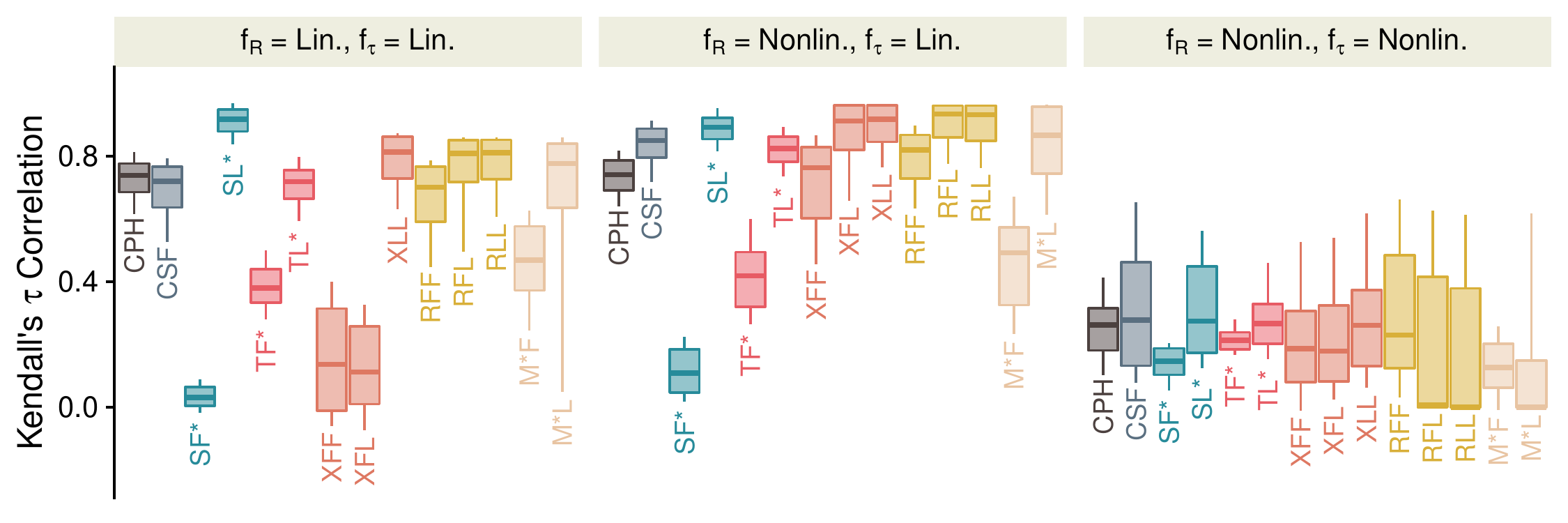}
\caption{Kendall's $\tau$ correlations of metalearners under unbalanced treatment assignment. DGPs and estimators are the same as in Figure \ref{fig:unbal}.}
\label{fig:unbal_corr}
\end{figure}

%\clearpage

\section{Additional Data Analysis Results}
\subsection{Additional Results from Global Null Analyses}

\input{Tables/Table6.tex}

\begin{figure}[ht!]
\centering
\includegraphics[width=0.9\textwidth]{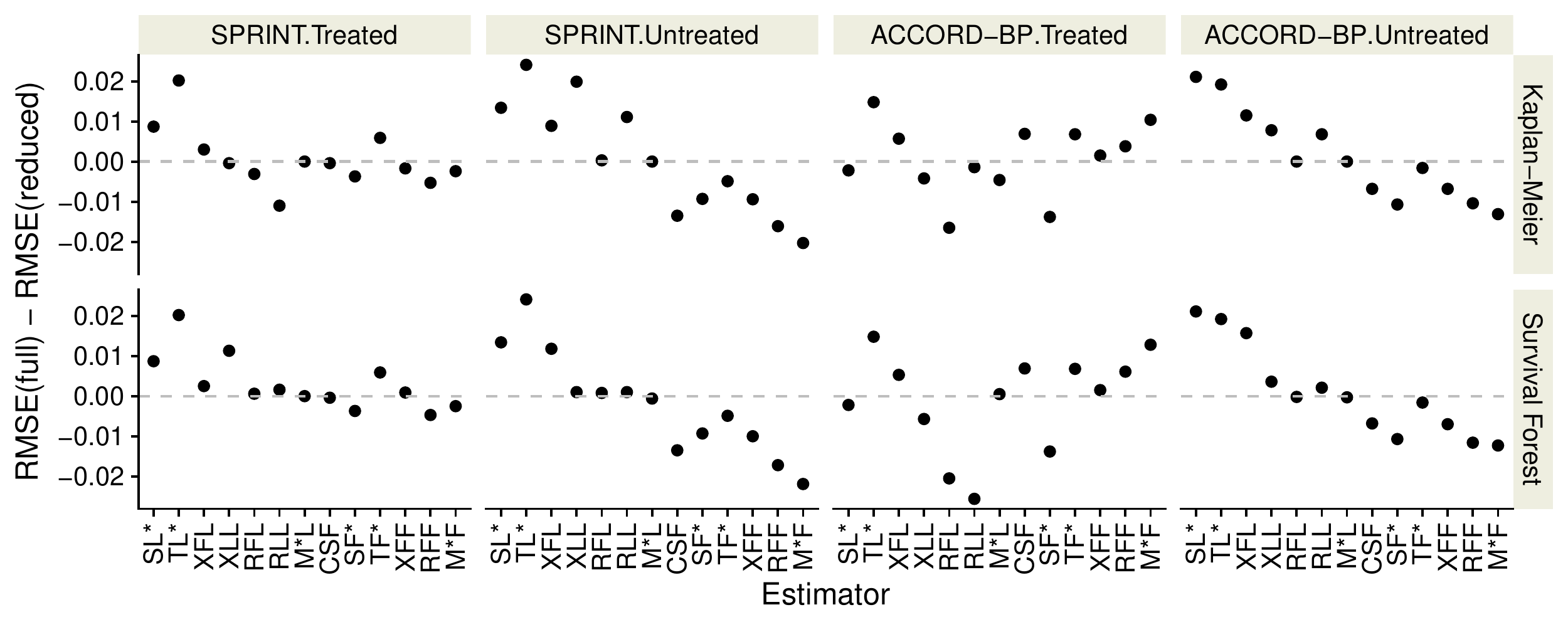}
\caption{Difference in root mean squared errors of metalearners under global null analysis. The difference is computed as the RMSE when all 13 original covariates (\emph{full}) are used for estimation minus the RMSE when only the PCE score and sub CKD are used as predictors (\emph{reduced}).}\vspace{-1em}
\label{fig:sprint_NULL_rmse}
\end{figure}
\vspace{1em}

\label{appendix:NULL}
\begin{figure}[H]
\centering
\includegraphics[width=0.9\textwidth]{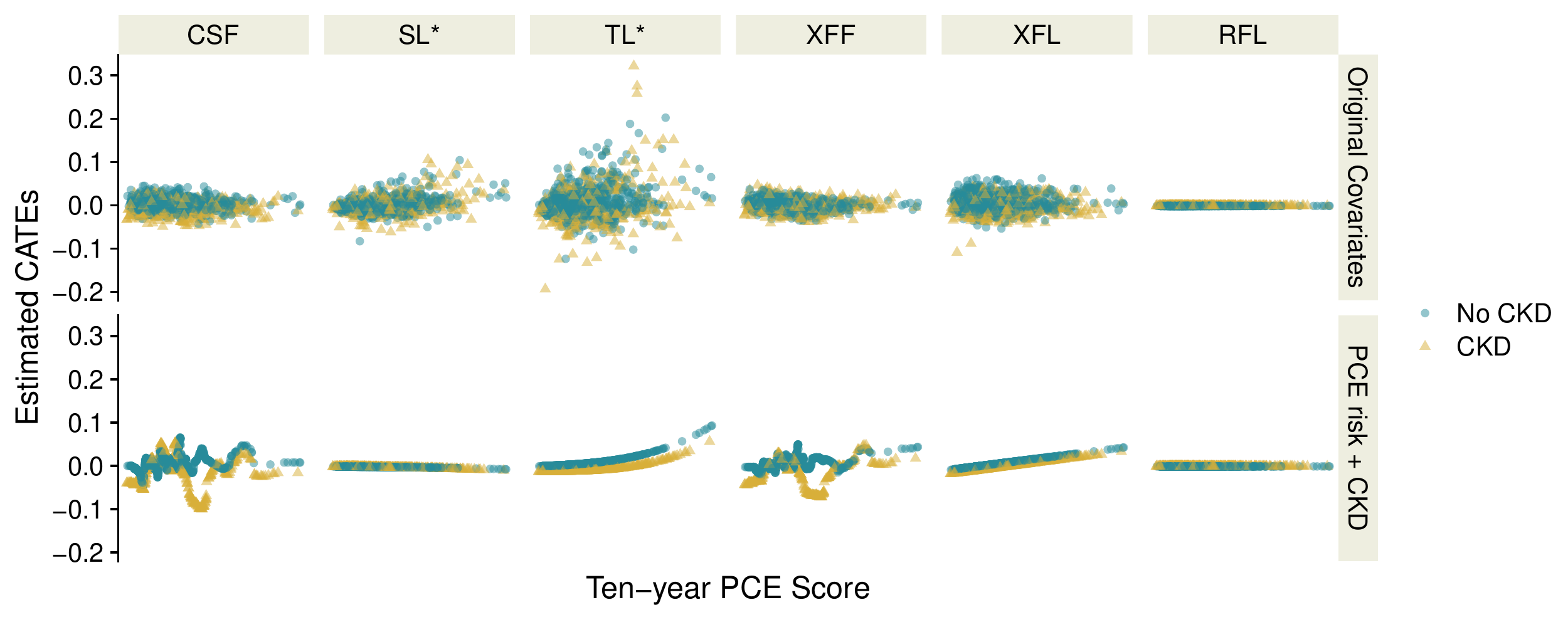}
\caption{Trends between ten-year PCE score and estimated CATEs in \emph{SPRINT}. The same analysis as in Figure \ref{fig:sprint_NULL} but used the \emph{Kaplan-Meier} estimator for modeling censoring instead.}\vspace{-1em}
\label{fig:sprint_NULL_ctl_KM}
\end{figure}

\begin{figure}[H]
\centering
\includegraphics[width=0.9\textwidth]{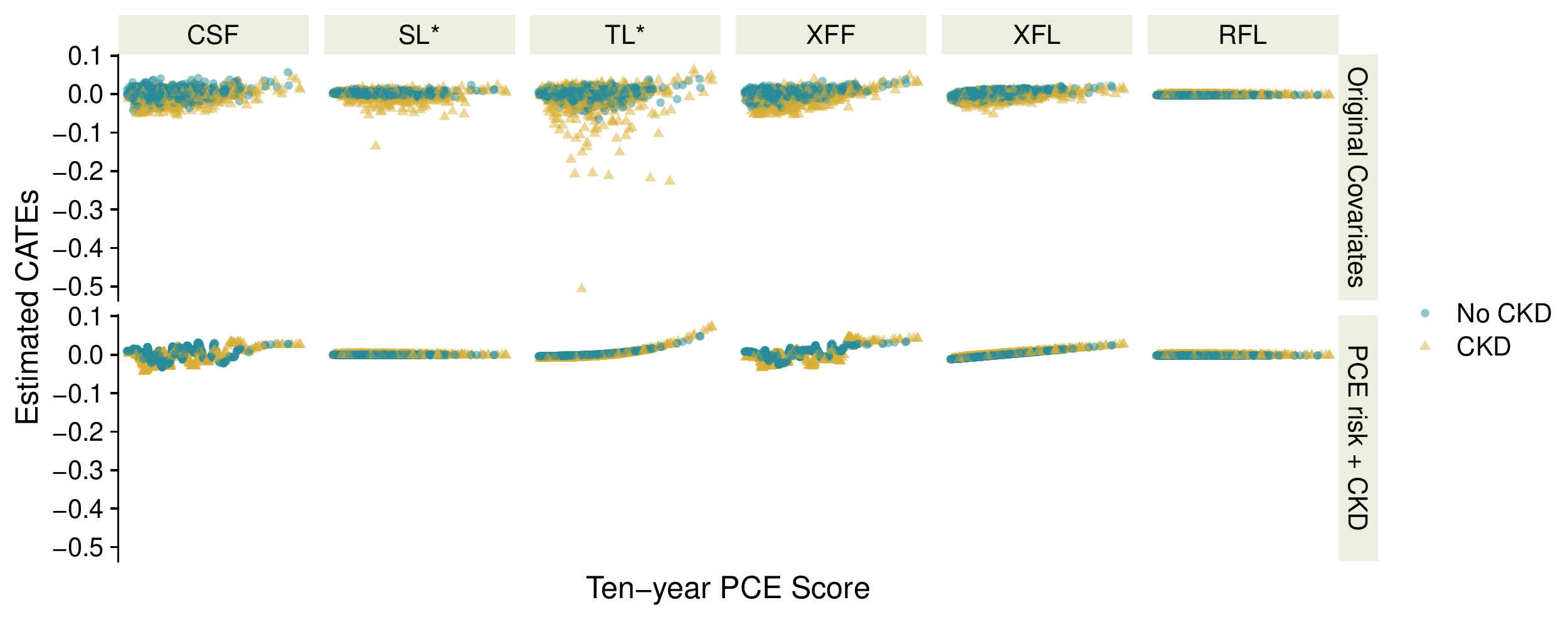}
\caption{Trends between ten-year PCE score and estimated CATEs in \emph{SPRINT}. The same analysis as in Figure \ref{fig:sprint_NULL} but conducted on \emph{treated} subjects and used a \emph{random survival forest} for modeling censoring instead.}
\label{fig:sprint_NULL_trt_SF}
\end{figure}

\begin{figure}[H]
\centering
\includegraphics[width=0.9\textwidth]{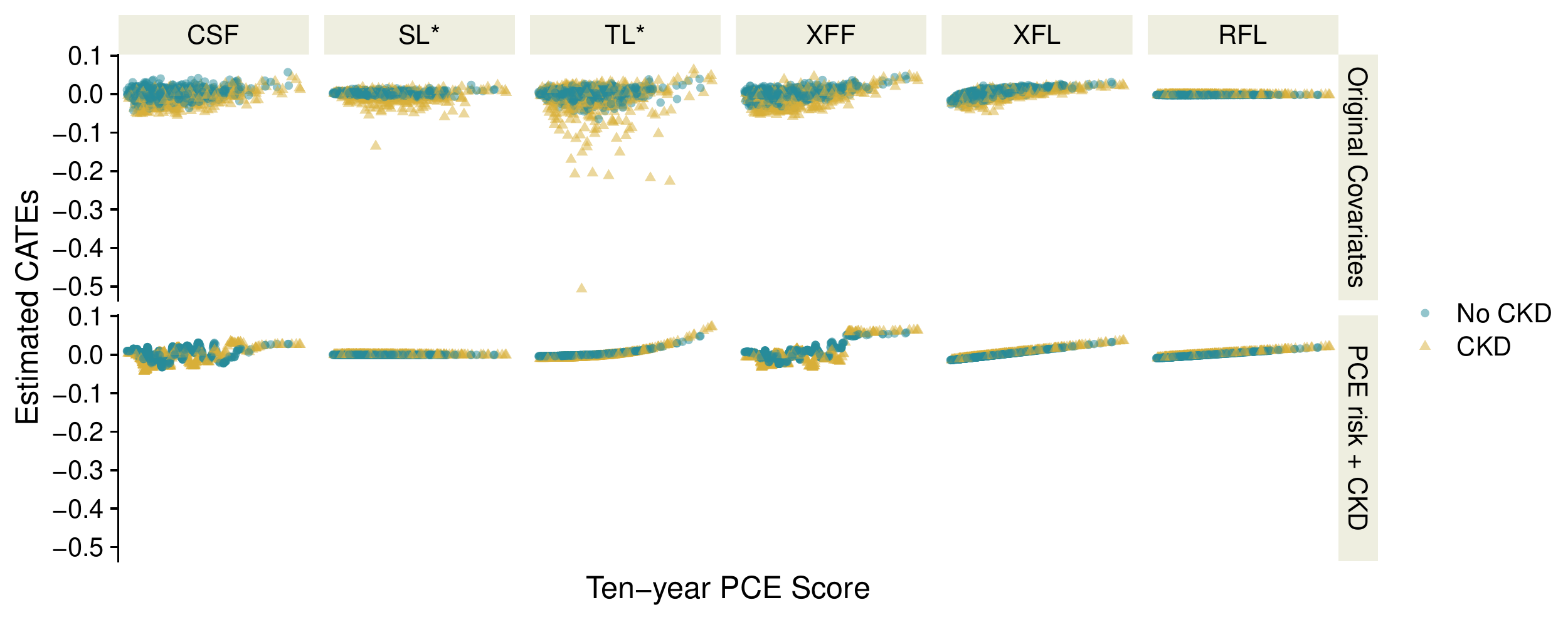}
\caption{Trends between ten-year PCE score and estimated CATEs in \emph{SPRINT}. The same analysis as in Figure \ref{fig:sprint_NULL} but conducted on \emph{treated} subjects and used the \emph{Kaplan-Meier} estimator for modeling censoring instead.}
\label{fig:sprint_NULL_trt_KM}
\end{figure}

\begin{figure}[H]
\centering
\includegraphics[width=0.9\textwidth]{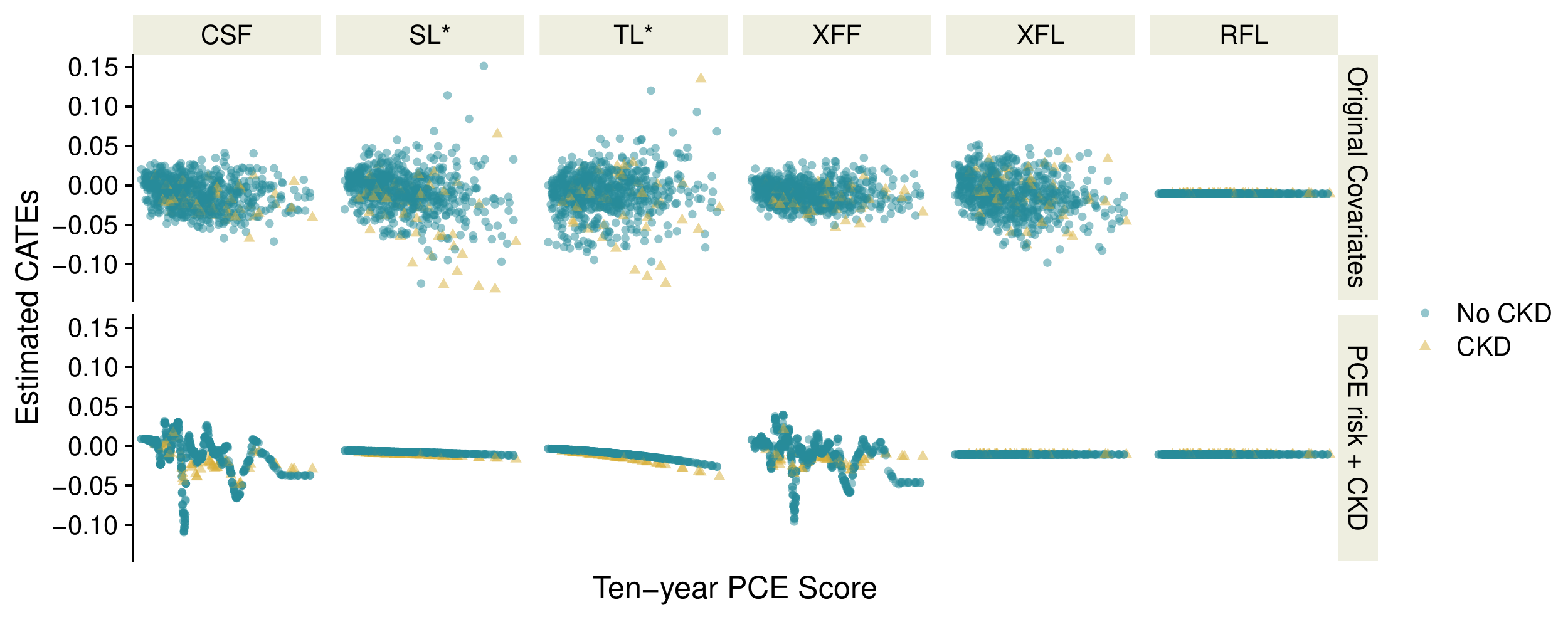}
\caption{Trends between ten-year PCE score and estimated CATEs in \emph{ACCORD}. The same analysis procedures in Figure \ref{fig:sprint_NULL} are used.}
\label{fig:accord_NULL_ctl_SF}
\end{figure}

\begin{figure}[H]
\centering
\includegraphics[width=0.9\textwidth]{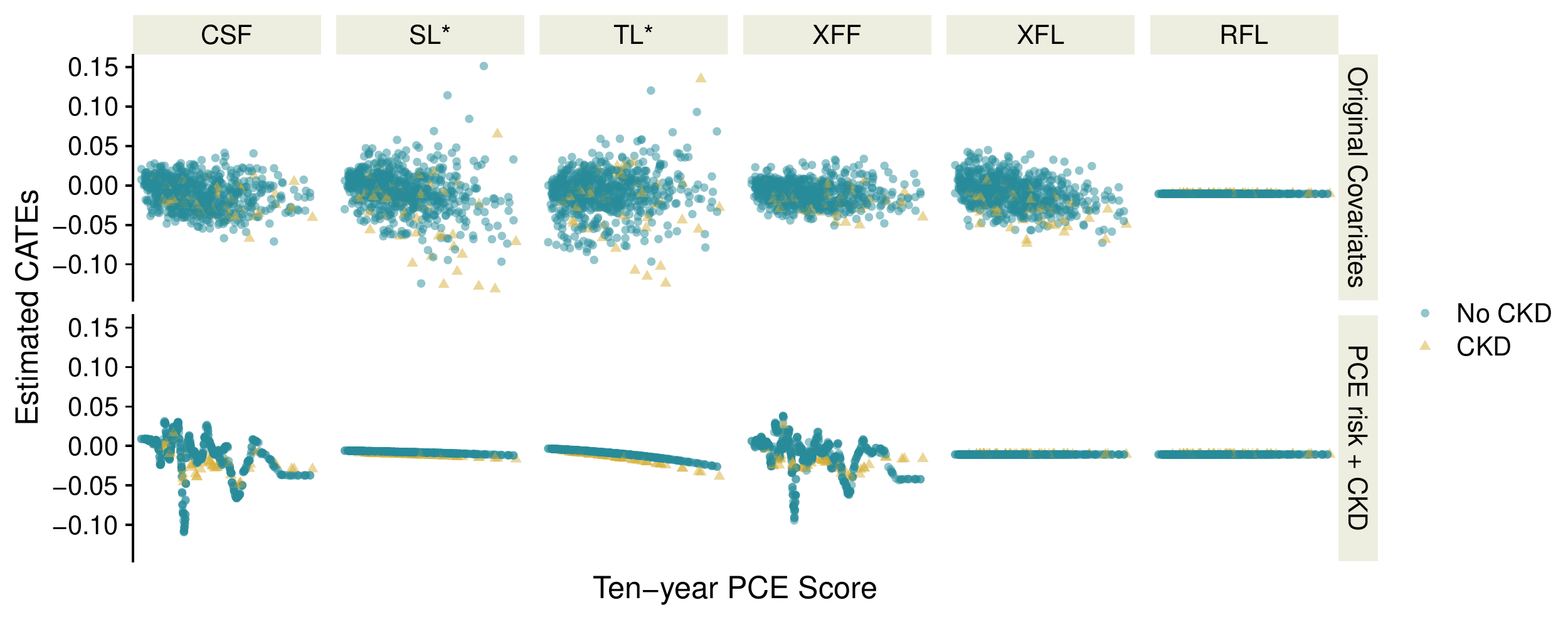}
\caption{Trends between ten-year PCE score and estimated CATEs in \emph{ACCORD}. The same analysis as in Figure \ref{fig:accord_NULL_ctl_SF} but used the \emph{Kaplan-Meier} estimator for modeling censoring instead.}
\label{fig:accord_NULL_ctl_KM}
\end{figure}

\begin{figure}[H]
\centering
\includegraphics[width=0.9\textwidth]{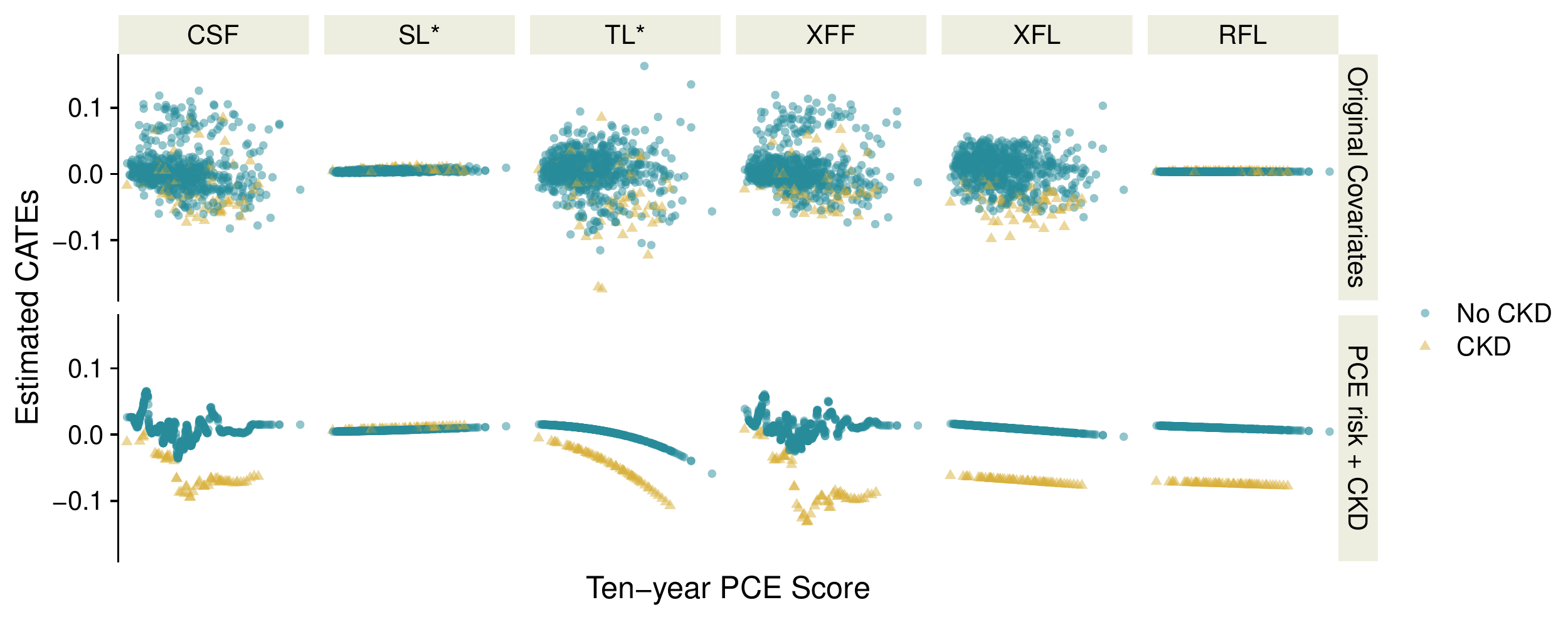}
\caption{Trends between ten-year PCE score and estimated CATEs in \emph{ACCORD}. The same analysis as in Figure \ref{fig:accord_NULL_ctl_SF} but conducted on \emph{treated} subjects and used a \emph{random survival forest} for modeling censoring instead.}\vspace{-1em}
\label{fig:accord_NULL_trt_SF}
\end{figure}

\begin{figure}[H]
\centering
\includegraphics[width=0.9\textwidth]{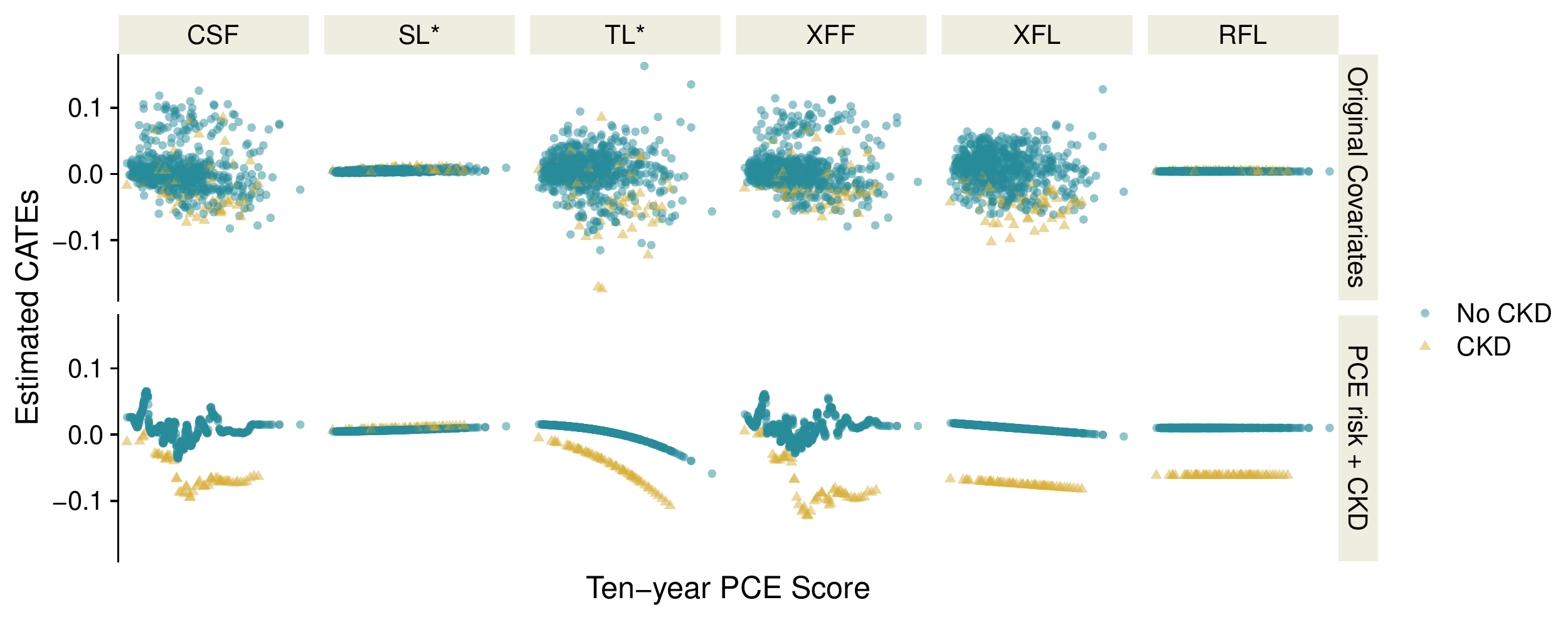}
\caption{Trends between ten-year PCE score and estimated CATEs in \emph{ACCORD}. The same analysis as in Figure \ref{fig:accord_NULL_ctl_SF} but conducted on \emph{treated} subjects and used the \emph{Kaplan-Meier estimator} for modeling censoring instead.}
\label{fig:accord_NULL_trt_KM}
\end{figure}

\subsection{Additional Figure from CATE Estimation in SPRINT and ACCORD}
\begin{figure}[H]
\centering
\includegraphics[width=0.9\textwidth]{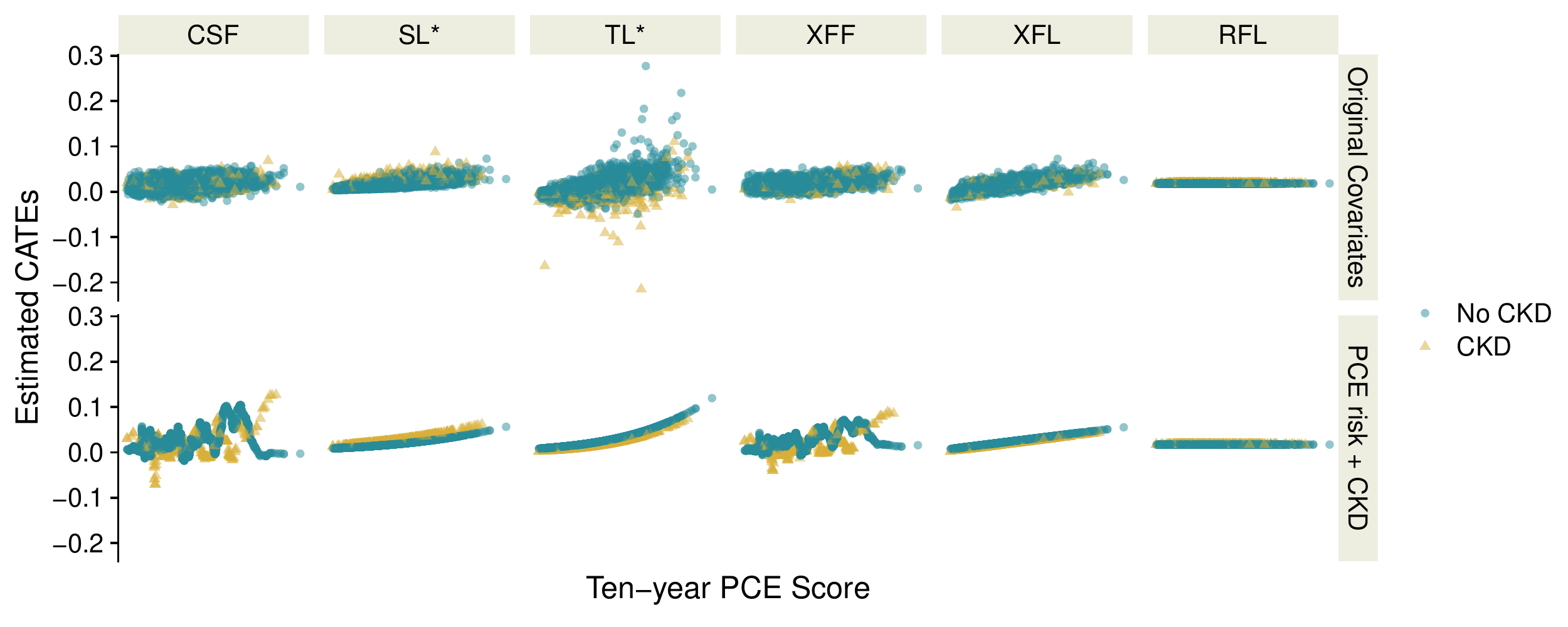}
\caption{Trends between ten-year PCE score and estimated CATEs in \emph{ACCORD-BP}. The same analysis procedures in Figure \ref{fig:sprint_full} are used.}
\label{fig:accord_full}
\end{figure}

\end{appendices}

\bibliographystyle{abbrvnat}
\bibliography{references}

\end{document}

%% file: Tables/Table4.tex
\begin{table}[ht]
\centering
\caption{Performance assessment of metalearners in CATE estimation via global null analysis. By construction, the true CATEs in the modified SPRINT and ACCORD datasets are zero for all subjects. Original baseline covariates are used. The metalearners showed various amounts of overestimation of treatment heterogeneity (in terms of RMSEs), which may lead to false discoveries. The analysis was repeated for subjects under the standard and intensive arms and replicated to model censoring using a random survival forest and the Kaplan-Meier estimator, separately.}
\label{t:zerocatefull}
\begin{tabular}{cccccccc}
  \hline
   \multicolumn{4}{c}{Standard} & \multicolumn{4}{c}{Intensive}\\ \hline
   \multicolumn{2}{c}{Survival Forest} & \multicolumn{2}{c}{Kaplan-Meier}&\multicolumn{2}{c}{Survival Forest} & \multicolumn{2}{c}{Kaplan-Meier}\\ \hline
   Estimator & RMSE & Estimator & RMSE & Estimator & RMSE & Estimator & RMSE \\ 
  \hline
    \multicolumn{8}{c}{SPRINT}\\ \hline
 SF* & 0.0005 & SF* & 0.0005 & SF* & 0.0003 & SF* & 0.0003 \\ 
  M*L & 0.0008 & M*L & 0.0007 & M*L & 0.0021 & M*L & 0.0017 \\ 
  RFL & 0.0011 & RFL & 0.0015 & RLL & 0.0024 & RFL & 0.0022 \\ 
  RLL & 0.0021 & RLL & 0.0112 & RFL & 0.0027 & RLL & 0.0022 \\ 
  XLL & 0.0085 & XFF & 0.0120 & SL* & 0.0090 & XLL & 0.0088 \\ 
  XFF & 0.0114 & RFF & 0.0123 & XFL & 0.0094 & SL* & 0.0090 \\ 
  RFF & 0.0118 & TF* & 0.0139 & XLL & 0.0127 & XFL & 0.0116 \\ 
  TF* & 0.0139 & CSF & 0.0145 & RFF & 0.0136 & RFF & 0.0137 \\ 
  M*F & 0.0141 & M*F & 0.0145 & TF* & 0.0141 & TF* & 0.0141 \\ 
  CSF & 0.0145 & SL* & 0.0161 & XFF & 0.0146 & CSF & 0.0151 \\ 
  SL* & 0.0161 & XFL & 0.0179 & M*F & 0.0147 & XFF & 0.0152 \\ 
  XFL & 0.0176 & XLL & 0.0205 & CSF & 0.0151 & M*F & 0.0152 \\ 
  TL* & 0.0351 & TL* & 0.0351 & TL* & 0.0280 & TL* & 0.0280 \\ \hline
   \multicolumn{8}{c}{ACCORD-BP}\\ \hline
  SF* & 0.0010 & SF* & 0.0010 & SF* & 0.0013 & SF* & 0.0013 \\ 
  M*L & 0.0103 & M*L & 0.0104 & RLL & 0.0026 & RLL & 0.0027 \\ 
  RFL & 0.0106 & RFL & 0.0108 & RFL & 0.0034 & RFL & 0.0038 \\ 
  RLL & 0.0121 & RLL & 0.0167 & SL* & 0.0047 & SL* & 0.0047 \\ 
  XLL & 0.0132 & XFF & 0.0177 & M*L & 0.0062 & M*L & 0.0065 \\ 
  XFF & 0.0178 & XLL & 0.0189 & XLL & 0.0144 & XLL & 0.0157 \\ 
  TF* & 0.0204 & TF* & 0.0204 & XFL & 0.0283 & XFL & 0.0305 \\ 
  RFF & 0.0212 & RFF & 0.0213 & XFF & 0.0329 & XFF & 0.0320 \\ 
  CSF & 0.0220 & CSF & 0.0220 & CSF & 0.0336 & CSF & 0.0336 \\ 
  M*F & 0.0253 & XFL & 0.0223 & TL* & 0.0338 & TL* & 0.0338 \\ 
  XFL & 0.0265 & M*F & 0.0245 & RFF & 0.0347 & RFF & 0.0346 \\ 
  SL* & 0.0291 & SL* & 0.0291 & TF* & 0.0367 & TF* & 0.0367 \\ 
  TL* & 0.0296 & TL* & 0.0296 & M*F & 0.0438 & M*F & 0.0421 \\ 
   \hline
 \end{tabular}
\end{table}

%% file: Tables/Table5.tex
\begin{table}[ht]
\centering
\caption{Internal and external validation performance of CATE estimation in SPRINT and ACCORD-BP, respectively. For internal validation, none of the five metalearners showed significant AUTOCs, suggesting the lack of treatment heterogeneity of intensive BP therapy. For internal validation, with the original covariates, CSF and XFF showed (marginally) significant AUTOCs. RECETH is the square root of the default calibration error given by ECETH.}
\label{t:sprint_eval}
\resizebox{\textwidth}{!}{\begin{tabular}{lcccc}
 \hline 
       & \multicolumn{2}{c}{Original covariates} & \multicolumn{2}{c}{PCE + CKD}\\ \hline
Method & AUTOC & RECETH & AUTOC & RECETH \\ \hline
  \multicolumn{5}{c}{Internal Validation}\\ \hline
  SL* & -0.0043 (-0.0260, 0.0174)  & 0.0211 (0.0000, 0.0452) & -0.0002 (-0.0223, 0.0218) & 0.0006 (0.0000, 0.0358) \\ 
  TL* & -0.0002 (-0.0206, 0.0202)  & 0.0023 (0.0000, 0.0404) &  0.0002 (-0.0209, 0.0212) & 0.0142 (0.0000, 0.0431) \\ 
  XFF &  0.0048 (-0.0162, 0.0258)  & 0.0135 (0.0000, 0.0386) & -0.0144 (-0.0348, 0.0059) & 0.0186 (0.0000, 0.0426) \\ 
  XFL & -0.0053 (-0.0270, 0.0165)  & 0.0235 (0.0000, 0.0469) & -0.0019 (-0.0232, 0.0195) & 0.0000 (0.0000, 0.0285) \\ 
  RFL &  0.0000 ( 0.0000, 0.0000)  & 0.0142 (0.0000, 0.0307) &  0.0044 (-0.0161, 0.0248) & 0.0000 (0.0000, 0.0301) \\ 
  CSF &  0.0012 (-0.0187, 0.0211)  & 0.0184 (0.0000, 0.0425) & -0.0172 (-0.0371, 0.0028) & 0.0451 (0.0185, 0.0644) \\  \hline
\multicolumn{5}{c}{External Validation}\\ \hline
  SL* & -0.0092 (-0.0281, 0.0097)   & 0.0077 (0.0000, 0.0332) & -0.0031 (-0.0201, 0.0138) & 0.0000 (0.0000, 0.0277) \\ 
  TL* & -0.0100 (-0.0274, 0.0073)   & 0.0242 (0.0000, 0.0421) & -0.0095 (-0.0261, 0.0072) & 0.0215 (0.0000, 0.0406) \\ 
  XFF &  0.0169 (-0.0006, 0.0345)   & 0.0000 (0.0000, 0.0201) & -0.0088 (-0.0261, 0.0085) & 0.0362 (0.0111, 0.0526) \\ 
  XFL & -0.0081 (-0.0254, 0.0091)   & 0.0201 (0.0000, 0.0391) & -0.0095 (-0.0261, 0.0071) & 0.0162 (0.0000, 0.0377) \\ 
  RFL &  0.0000 ( 0.0000, 0.0000)   & 0.0087 (0.0000, 0.0229) &  0.0000 ( 0.0000, 0.0000) & 0.0074 (0.0000, 0.0221) \\ 
  CSF &  0.0178 ( 0.0010, 0.0345)   & 0.0000 (0.0000, 0.0294) & -0.0111 (-0.0282, 0.0060) & 0.0321 (0.0000, 0.0486) \\   \hline
\end{tabular}}
\end{table}

%% file: Tables/Table6.tex
\begin{table}[ht]
\centering
\caption{Performance assessment of metalearners in CATE estimation via global null analysis. By construction, the true CATEs in the modified SPRINT and ACCORD datasets are zero for all subjects. The estimated ten-year CVD risk (using the pooled cohort question) and subclinical CKD are used as predictors. The metalearners showed various amounts of overestimation of treatment heterogeneity (in terms of RMSEs), which may lead to false discoveries. The analysis was repeated for subjects under the standard and intensive arms and replicated to model censoring using a random survival forest and the Kaplan-Meier estimator, separately.}
\label{t:zerocatepce}
\begin{tabular}{cccccccc}
  \hline
   \multicolumn{4}{c}{Standard} & \multicolumn{4}{c}{Intensive}\\ \hline
   \multicolumn{2}{c}{Survival Forest} & \multicolumn{2}{c}{Kaplan-Meier}&\multicolumn{2}{c}{Survival M*orest} & \multicolumn{2}{c}{Kaplan-Meier}\\ \hline
   Estimator & RMSE & Estimator & RMSE & Estimator & RMSE & Estimator & RMSE \\ 
  \hline
    \multicolumn{8}{c}{SPRINT}\\ \hline
  RFL & 0.0001 & RLL & 0.0001 & SL* & 0.0003 & SL* & 0.0003 \\ 
  RLL & 0.0011 & XLL & 0.0006 & RLL & 0.0008 & M*L & 0.0017 \\ 
  M*L & 0.0014 & M*L & 0.0007 & XLL & 0.0014 & SF* & 0.0040 \\ 
  SL* & 0.0027 & RFL & 0.0012 & RFL & 0.0020 & RFL & 0.0053 \\ 
  XFL & 0.0058 & SL* & 0.0027 & M*L & 0.0021 & TL* & 0.0078 \\ 
  XLL & 0.0075 & XFL & 0.0090 & SF* & 0.0040 & TF* & 0.0082 \\ 
  SF* & 0.0098 & SF* & 0.0098 & XFL & 0.0069 & XFL & 0.0086 \\ 
  TL* & 0.0110 & TL* & 0.0110 & TL* & 0.0078 & XLL & 0.0092 \\ 
  TF* & 0.0188 & TF* & 0.0188 & TF* & 0.0082 & RLL & 0.0132 \\ 
  XFF & 0.0214 & XFF & 0.0214 & XFF & 0.0137 & CSF & 0.0155 \\ 
  RFF & 0.0249 & RFF & 0.0243 & CSF & 0.0155 & XFF & 0.0169 \\ 
  CSF & 0.0280 & CSF & 0.0280 & RFF & 0.0167 & M*F & 0.0176 \\ 
  M*F & 0.0360 & M*F & 0.0348 & M*F & 0.0172 & RFF & 0.0187 \\\hline
   \multicolumn{8}{c}{ACCORD-BP}\\ \hline
  SL* & 0.0080 & SL* & 0.0080 & M*L & 0.0057 & RLL & 0.0041 \\ 
  XLL & 0.0096 & RLL & 0.0099 & SL* & 0.0069 & SL* & 0.0069 \\ 
  RLL & 0.0100 & TL* & 0.0104 & SF* & 0.0151 & M*L & 0.0111 \\ 
  TL* & 0.0104 & M*L & 0.0104 & TL* & 0.0190 & SF* & 0.0151 \\ 
  M*L & 0.0106 & RFL & 0.0107 & XLL & 0.0201 & TL* & 0.0190 \\ 
  RFL & 0.0107 & XFL & 0.0108 & XFL & 0.0230 & XLL & 0.0199 \\ 
  XFL & 0.0108 & XLL & 0.0111 & RFL & 0.0243 & RFL & 0.0205 \\ 
  SF* & 0.0117 & SF* & 0.0117 & CSF & 0.0267 & XFL & 0.0248 \\ 
  TF* & 0.0220 & TF* & 0.0220 & RFF & 0.0267 & CSF & 0.0267 \\ 
  XFF & 0.0248 & XFF & 0.0245 & RLL & 0.0282 & RFF & 0.0277 \\ 
  RFF & 0.0287 & CSF & 0.0288 & TF* & 0.0299 & TF* & 0.0299 \\ 
  CSF & 0.0288 & RFF & 0.0293 & M*F & 0.0310 & XFF & 0.0305 \\ 
  M*F & 0.0376 & M*F & 0.0376 & XFF & 0.0314 & M*F & 0.0317 \\ 
   \hline
 \end{tabular}
\end{table}